\documentclass[global,twocolumn]{svjour}

\usepackage{fancyhdr}
\usepackage{fullpage}
\usepackage{amsmath}
\usepackage{amsbsy}
\usepackage{amssymb}
\usepackage{amscd}
\usepackage{amsfonts}
\usepackage{supertabular}
\usepackage{graphics}
\usepackage{graphicx}
\usepackage{verbatim}
\usepackage{subfigure}
\usepackage{epsfig}
\usepackage{xspace}
\usepackage{euscript}
\usepackage{alltt}
\usepackage{boxedminipage}
\usepackage{float}
\usepackage{times}
\usepackage[colorlinks]{hyperref}
\usepackage[square,authoryear]{natbib}

\newcommand{\mbs}[1]{\boldsymbol{#1}}

  \def\bx{{\mbs{x}}}

\makeatletter \@addtoreset{figure}{section}
\def\thefigure{\thesection.\@arabic\c@figure} \def\fps@figure{h, t}
\@addtoreset{equation}{section}
 \makeatother

\begin{document}

\title{Optimal design of composite granular protectors}
\author{Fernando~Fraternali\inst{1}
\and Mason~A.~Porter\inst{2} \and Chiara~Daraio\inst{3}} 
\institute{Department of Civil
Engineering, University
  of Salerno, 84084 Fisciano (SA), Italy \and Oxford Centre for Industrial and Applied Mathematics, Mathematical Institute, University of Oxford, OX1 3LB, United Kingdom \and Graduate Aeronautical Laboratories (GALCIT) and Department of Applied Physics,  \\ 
California Institute of Technology, Pasadena, CA 91125, USA. Corresponding author. Email: daraio@caltech.edu} \date{Received: \today / Revised version: \today}

\maketitle
\begin{abstract}

We employ an evolutionary algorithm to investigate the optimal design of composite protectors using one-dimensional granular chains composed of beads of various sizes, masses, and stiffnesses.  We define a fitness function using the maximum force transmitted from the protector to a ``wall" that represents the body to be protected and accordingly optimize the {topology} (arrangement), {size}, and {material} of the chain.  We obtain optimally randomized granular protectors characterized by high-energy equipartition and the transformation of incident waves into interacting solitary pulses.  We consistently observe that the pulses traveling to the wall combine to form an extended (long-wavelength), small-amplitude pulse.  
\end{abstract}
%

\section{Introduction}

One-dimensional (1D) lattices (chains) of particles interacting according to nonlinear potentials have been receiving increasing attention in the scientific community because of their special wave dynamics, which allows energy transport through solitary waves \cite{mackay99, nesterenko1,senmanciu01, HeatPhysicsReport,focus,friep1,friema}. In the case of granular systems, particle interactions are strongly nonlinear because of nonlinear compressive forces and no-tension behavior \cite{nesterenko1,senmanciu01,wang,hinch99,sen08}.  As a result, granular lattices can support traveling solitary waves with compact support \cite{nesterenko1}.  The evolution of nonlinear particle systems toward energy equipartition (or \textit{thermalization}), predicted by statistical mechanics \cite{pathria}, is also particularly interesting.  Stable or transient energy transport through coherent modes (solitary waves, breathers, etc.) can develop (\cite{friep1, friep2, friep3, friep4, flach, peyrard, mirnov, eleft}), and eventual thermalization might not occur at all, as investigated in great detail for the Fermi-Pasta-Ulam problem and related nonlinear lattice systems \cite{focus}.

By designing protectors or containers optimally, the strongly nonlinear dynamics of granular systems can be exploited to produce fast decomposition of an external impulse into trains of solitary waves, energy trapping, shock disintegration, and more \cite{hong05,nesterenko2,dar05b,dar06,dimer,donsen05,donsen06,wang}.  Furthermore, it has been emphasized that using a suitable randomization of the granular system--involving, for example, the variation of particle sizes, masses, and materials--one might induce nonuniformity in the steady states in the velocity profiles; the appearance of negative velocities; marked thermalization; wave-amplitude decay; and anomalous features of wave propagation through interfaces between particles differing in masses, sizes, and/or mechanical properties \cite{nesterenko1,nesterenko2,dar05b,hong05,job}.  

When dealing with the optimal design of granular protectors, one can optimize features such as particle distribution, connectivity, size, and material through either discrete or continuous approaches.  (These ideas are known, respectively as {\it shape}, {\it topology}, {\it size}, and {\it material} optimization.)  Discrete approaches introduce suitable \textit{ground structures}--which refer to background structures in which the material densities of predefined connections are subject to optimization \cite{rin85,roz93,ped93,ben03,kir96}.  Continuous models instead use \textit{homogenization theory}, as they examine design domains with perforated composite microstructures \cite{ben88,ben89,kik95,jac98,all01,ben03}.  Well-established gradient-based optimization techniques include {mathematical programming}, {optimality criteria} (that is, suitable mathematical conditions defining an optimally designed structure) \cite{sav85,ber87}, {sequential approximate optimization} \cite{sva87,fle87,ma95}.  Available methods that are not based on gradients include {simulated annealing} \cite{bal91,shi97}, {biological growth} \cite{mat90,mro03}, and {genetic and evolutionary algorithms} \cite{gol86,jen92,cha94,raj95,kic05}.

{\it Evolutionary Algorithms} (EAs) provide a family of optimization methods inspired by Darwin's theory of evolution. They search for the best ``phenotype" in a given population of candidate solutions by applying selection mechanisms and genetic operators similar to the intermingling of chromosomes in cell reproduction and replication.  First, one evaluates each element (individual) of the population in terms of a quantitative fitness, which represents the feature that discriminates between phenotypes.  One then mates the individuals using a recombination operator.  Finally, one mutates a given percentage of the individuals, thereby creating a new population.  One then repeats these steps cyclically until some termination criterion is reached.  The individual with the best fitness in the final population provides a guess of the global optimum of the fitness function.  EAs are a natural fit for optimization problems in granular systems, as in such problems one can easily identify the system particles (i.e., the beads) with cells and their geometrical and mechanical properties (including radii, mass densities, elastic moduli, material types, and so on) with the corresponding genes. Furthermore, EAs  require little knowledge of the search environment, can escape from local optima (in order to achieve a better global optimum), and are well-suited to problems with large and complex solution spaces \cite{kic05} (such as those arising from the optimization of strongly nonlinear dynamical systems).

The present work exploits EAs for the optimal design of composite granular protectors. We identify the fitness function with the  force $F_{out}$ transmitted from the protector to a ``wall" that represents the body to be protected.  We compute the performance of the candidate solutions under given impact loadings through a Runge-Kutta time-discretization of Hamilton's equations of motion.  We adopt the hard-sphere model of interactions between adjacent beads for computational reasons, as a very large number of simulations are required by the optimization process.  We also ignore dissipative effects, in accord with the standard models in the literature \cite{nesterenko1}.  We note, however, that including relevant dissipative effects \cite{rosas03,rosas07} such as friction, plasticity, large deformations, and so on, using (for example) time-stepping techniques of non-smooth contact dynamics \cite{moreau} or molecular dynamics \cite{herrmann}, would not change the above EA framework. Dissipation is expected to enhance the effectiveness of the protector by further reducing the force amplitude transmitted at the wall, as shown (for example) in the experimental results reported below.

In this paper, we investigate several optimization problems.  We focus, in particular, on topology, size, and material optimization of 1D composite granular chains.  We compare the dynamics of the optimized systems we obtain with those of granular protectors and special granular systems (\textit{sonic vacua}) available in the literature \cite{nesterenko1,donsen06,dar06,hong05}. We show that the use of EAs offers a dramatic advantage in the design of granular protectors, leading to a significant decrease of the transmitted force. This EA-driven optimal design generates suitable topology, size, and material randomization by combining effects of wave disintegration and reflection at the interfaces between geometrical and/or mechanical discontinuities.  A general feature we observe in the optimized protectors is the transformation of incident waves into a collection of interacting reflected and transmitted solitary pulses, which in particular form an extended (long-wavelength), small-amplitude wave that travels to the wall.  We also find that optimization randomizes these systems (adding to their disorder) and produces a marked thermalization.  We constantly observe (in the absence of forced symmetry constraints) the appearance of soft/light beads near the wall, hard/heavy beads near the end impacted by the striker, and alternating hard and soft beads in the central section of the optimized chains. The observed ``shock mitigation" behavior allows one to think of granular protectors in a new way--as tunable kinetic systems rather than as purely dissipative systems.  Consequently, they offer the exciting possibility of creating much more effective energy transformation and shielding devices. 

The remainder of the paper is organized as follows.  In Section \ref{mech}, we formulate our mechanical and numerical models.  We then show how to optimize granular protectors in Section \ref{granprot}.  We consider, in turn, topology optimization, size optimization, periodic sequences of optimized cells, and material optimization.  As an extended example, we investigate the optimization of a container proposed by Hong \cite{hong05} with both impulsive and shock-type loading.  Finally, we summarize our results in Section \ref{conclusions}.

\section{Mechanical and numerical modeling} \label{mech}

Consider a non-dissipative chain of $N$ granular particles described by the Hamiltonian \cite{nesterenko1}
\begin{align} \label{eq:Hamiltonian}
  	H \ = \ \sum_{i=1}^{N} \left( \frac{1}{2} \frac{p_i^2}{m_i} + V_i( q_{i}   - q_{i+1} ) - W_i(q_i) \right),
\end{align}
where $m_i$, $q_i$ and $p_i$, respectively, denote the mass, the displacement from the "packed"  configuration (particles touching each other without deformation) and the momentum of the $i$th particle,  $V_i$ is the potential of the interaction force between particles $i$ and $i+1$, and $W_i$ is the potential of the external forces acting on the $i$th particle (including gravity, static precompression, etc.).   We introduce an $(N+1)$st particle in order to model a {wall} that constrains the chain.  In so doing, we assume  that $p_{N+1} = q_{N+1}=0$ during the motion.  The Hamiltonian (\ref{eq:Hamiltonian}) yields a system of $2N$ first-order differential equations describing the motion of the system:
\begin{align}\label{eq:motioneqns}
 	\dot{p}_i \ = \ - \frac{\partial H}{\partial q_i}, \ \ \ \ \  \dot{q}_i \ = \ \frac{\partial H}{\partial p_i}, \ \ \ \ \  i=1,\cdots,N\,,
\end{align}
to be solved with the initial conditions $p_i(t=0)=p_i^{(0)}$,  $q_i(t=0)=q_i^{(0)}$, where $t \in [0,\bar{t}]$ denotes the time variable, $\bar{t}$ indicates the final observation instant, and a dot over a variable denotes its derivative with respect to time.  

Assuming that stresses remain within the elastic threshold and that particle contact areas and velocities are sufficiently small, we introduce tensionless, Hertzian type power-law interaction potentials \cite{nesterenko1}
\begin{align}\label{eq:Hertz}
	{V}_i \ \  = \ \  \frac{1}{n_i + 1} \ \alpha_i \ [ (q_{i}  - q_{i+1} )^+]^{n_i + 1}\,,
\end{align}
where $\alpha_i$ and $n_i$ are coefficients depending on material properties and particle geometry,  and $(\cdot)^+$ denotes the positive part of $(\cdot)$. Most of the examples that we examine in this paper are spherical grains, for which Hertz's law implies
\begin{align}\label{eq:alpha}
 	\alpha_i \ = \ \frac{4 {\cal{E}}_i {\cal{E}}_{i+1} \sqrt{\frac{r_i r_{i+1}}{r_i + r_{i+1}}}} {3 {\cal{E}}_{i+1}(1 - \nu_i^2) + 3 {\cal{E}}_i(1 - \nu_{i+1}^2)}, \ \ \  n_i = \frac{3}{2}\,,
\end{align}
where $r_i$, ${\cal{E}}_i$, and $\nu_i$ denote, respectively, the radius, elastic (Young) modulus, and Poisson ratio of particle $i$.  In the case of the granular container investigated by Hong \cite{hong05}, we instead use the values shown in Table~\ref{tab:Mat2} for $\alpha_i$ and $n_i$.

Additionally, let
\begin{align}\label{eq:TV}
  T =  \sum_{i=1}^{N} \frac{1}{2} \frac{p_i^2}{m_i}\,, \quad
  V  =  \sum_{i=1}^{N} \left[V_i(q_i - q_{i+1}) - W_i(q_i) \right]\,,
\end{align}
and ${E} = H = T+V$, where $T$ denotes the system's kinetic energy, $V$ denotes the potential energy, and ${E}$ denotes the total energy.  We also introduce the local energies 
\begin{align} \label{eq:Ei}
  	{E}_i \ = \ \frac{1}{2} \frac{p_i^2}{m_i} + \frac{1}{2} \left[ V_i(q_{i-1} - q_{i}) + V_i(q_{i} - q_{i+1}) \right]  \,
\end{align}
at each site (bead), and the \textit{energy correlation function} (which is slightly different from that introduced in Ref.~\cite{eleft})
\begin{align} \label{eq:corr1}
  	C(t,0) \ = \  \frac{c(t)}{c(0)}\,,
\end{align}
where 
\begin{align} \label{eq:corr2}
  	c(t) \ = \ \frac{1}{N}  \left\langle \sum_{i=1}^{N} {E}_i^2(t) \right\rangle \  \ - \  \
	                   \left\langle \frac{1}{N} \sum_{i=1}^{N} {E}_i(t)  \right\rangle^2 \,,
\end{align}
$\left\langle \cdot \right\rangle$ denotes the average over time (from $0$ to the current time), and $C(t,0)$ indicates how the energy is transferred between the different beads.  Observe that $C(t,0) = 0$ corresponds to energy equipartition. 

Equations (\ref{eq:motioneqns}) can be solved numerically using a standard fourth-order Runge-Kutta integration scheme (as discussed in, for example, Ref.~\cite{nesterenko1}) with a time integration step of
\begin{align}\label{eq:Dt}
	\Delta t \ \  = \ \ k \left[  \min_{i=1,...,N} \left\{ \frac{r_i}{c_i} \right \}\right]\,,
\end{align}
where $c_i$ is the sound speed in the material for the $i$th particle and $k \in (0,1]$ is a scaling  factor. Equation (\ref{eq:Dt}) gives a time-integration step of about $2 \times 10^{-8}$ s for $k=0.1$ and 1 mm stainless steel bead chains (see Table~\ref{tab:Mat1}), ensuring relative errors lower than $10^{-8}$ in the total energy conservation for times up to few thousand $\mu s$. 

We now assume that the configuration of the granular system is described by a collection of design variables or ``genes" (which can include particle radii, mass densities, elastic moduli, material types, etc.)
\begin{equation}\label{eq:p_def}
  	\bx =  \left\{x_i \right\}_{i=1,\ldots, M} \,,
\end{equation}
subject to simple bounds of the form
\begin{equation}\label{eq:x_bounds}
  \bx \in X
  =
  [ x^{lb}_1, x^{ub}_1 ] \times  \ldots \times [ x^{lb}_M, x^{ub}_M ]\,.
\end{equation}
One can always assume assume that $x^{lb}_i = 0$ and $ x^{ub}_i = 1$, for all $i \in \{1,...,M\}$ through suitable rescaling of design variables.

\begin{figure}[tbp]
    \centerline{\includegraphics[angle=90,width=80mm]{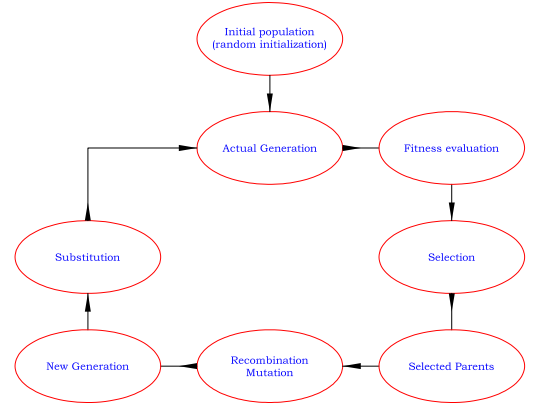}}
    \caption{\scriptsize{(Color online) Diagrammatic representation of an evolutionary algorithm.}}\label{Cycle}
\end{figure}

Given an assigned $\bx$, a numerical simulation of the system dynamics under a prescribed impulse or shock loading gives the protection performance (fitness) $f = \| F_{out} \|_{L^{\infty}}$ of the corresponding design configuration.  Here, $F_{out}$ denotes the force transmitted from the system to the wall, and $\| F_{out}\|_{L^{\infty}}$ denotes its norm with respect to the Sobolev space $L^{\infty}([0,\bar{t}])$ \cite{adams}. The optimal design configuration $\bx_{opt}$ can then be identified with the solution of the multivariate optimization problem,
\begin{equation}\label{eq:x_opt}
  	\min_{\bx \in X} f(\bx)\,,
\end{equation}
which is expected to be influenced by multiple local optima.  The problem (\ref{eq:x_opt}) can be conveniently solved via EAs (see, for example, Refs.~\cite{gol86,jen92,haj93,cha94,raj95,kic05}) through the cyclic iterative procedure illustrated in Fig.~\ref{Cycle}.  In the present paper, we will use the Breeder Genetic Algorithm (BGA) presented in Ref.~\cite{dcioppa96}.  BGAs, in contrast to other EAs (in which the selection is stochastic), selects only from among the $T_R\%$ best elements of the current population of $N$ individuals (where $T_R\%$ denotes the so-called \textit{truncation rate}) to be recombined and mutated (mimicking animal breeding).  This feature makes the BGAs more efficient than standard EAs for performing optimization in large search spaces \cite{Muh91,Muh94}.

\nobreak

\section{Optimization of granular protectors} \label{granprot}

We deal in the following sections with topology, size, and material optimization of 1D composite granular protectors subject to impulsive and shock-type loadings.  We employ formula (\ref{eq:Dt}) with $k=0.1$ for time discretization and always assume that genes are continuous variables ranging over $[0,1]$ with a population size of 50 individuals; an initial, randomly-chosen, truncation rate ($T_R$) equal to 15$\%$, Extended Intermediate Recombination (EIR) \cite{Muh91}, and a mutation rate in the interval $[10\%, 50\%]$.  (We use the value $10\%$ for size optimization, which consists of genuinely continuous genes and $50\%$ in all the other examples, which instead model discrete design variables using continuous genes.)  EIR generates offspring along the line defined by the parents in the search space and allows one to also create offspring outside of the segment joining the parents. See \cite{dcioppa96, Muh91, Muh94} for further technical details of the employed BGA.   

The examples of Sections~\ref{topology} and \ref{size} consider chains of stainless steel beads, whereas those in Section~\ref{material} examine a composite chain composed of polytetrafluoroethylene (PTFE) and stainless steel beads.  We show the material properties of these beads in Table~\ref{tab:Mat1}.  The final example, discussed in Section~\ref{hongs}, considers a long composite chain--the protector recently investigated by Hong \cite{hong05})--with the material properties shown in Table~\ref{tab:Mat2}. We studied protectors with one end in unilateral contact with a rigid wall (simulating the body to be protected) and the other end free. In most cases, we assumed that the free end was impacted by a striker; in the final example, we assumed that it was loaded by a prescribed force. We focused our attention on the short-term dynamics of the protector over an observation time slightly larger than that necessary to transmit the input actions to the wall.

\begin{table}[htbp]
  \begin{center}
    \begin{tabular}{|c|c|c|c|}
      \hline
      Label & $m$  & $\alpha$ &  $n$ \\
      \hline
       mat1 & $2.0$ & 5657 & 1.0 \\
      \hline
       mat2 & $1.0$ & 5657 & 2.0 \\
      \hline
       mat3 & $0.3$ & 5657 & 1.5 \\
      \hline
       mat4 & $0.1$ & 5657 & 1.5 \\
      \hline
       \end{tabular}
    \end{center}
    \caption{Material properties (mass $m$, contact coefficient $\alpha$, and contact exponent $n$) of the granular container investigated by Hong \cite{hong05} (in abstract units, as discussed in the main text).}
  \label{tab:Mat2}
\end{table}

\begin{table}[htbp]
  \begin{center}
    \begin{tabular}{|c|c|c|c|}
      \hline
      & $\rho$ ($\mbox{kg/m}^3$)  & $\cal{E}$ (GPa) & $\nu$  \\
      \hline
       stainless steel & 8000  & 193.00 & 0.30  \\
      \hline
       PTFE & 2200 & 1.46 &0.46 \\
      \hline
        \end{tabular}
    \end{center}
    \caption{Material properties (mass density $\rho$, elastic modulus ${\cal{E}}$, and Poisson ratio $\nu$) of stainless steel and PTFE beads.}
  \label{tab:Mat1}
\end{table}

\subsection{Topology optimization}\label{topology}

\begin{figure}[tbp]
    \centerline{\includegraphics[angle=0,width=85mm]{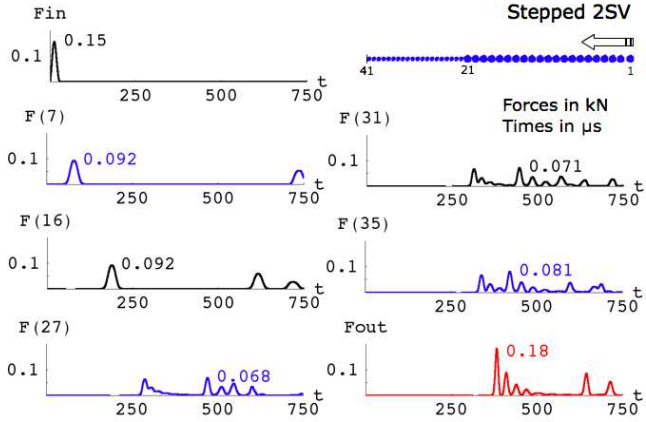}}
    \caption{\scriptsize{(Color online) Force versus time plots for the stepped two sonic vacua (2SV).  The striker impacts the end with larger-radius beads.    
      }}\label{2SV_force}
\end{figure}

Nesterenko used the monicker \textit{sonic vacua} to describe an unprecompressed (or weakly-precompressed) granular chains because the sound speed is zero or very small in such systems \cite{nesterenko1}. He studied the behavior of two adjacent monodisperse sonic vacua (2SV), characterized by a sharp variation in bead size (a \textit{stepped} 2SV), under the impact of a striker.  He observed two remarkable phenomena: disintegration of the incident pulse into a solitary wave train when it passes from the sub-chain with larger radius to the one with smaller radius; and a partial reflection in the opposite case (see also Ref.~\cite{job}).  Here we examine the topology optimization of a stepped 2SV in order to determine the particle arrangement that minimizes $F_{out}$ under a given impact event.  Figure \ref{2SV_force} shows different force-time histories in a 2SV hit by a striker at the sub-chain with larger radius.  The system is composed of 20 large beads of radius $r = r_L=3.95$ mm, 20 small beads of radius $r =r_S= 2.375$ mm, and the striker (particle number 1), which has radius $r=r_L$ and initial velocity $v = 1$ m/s (see $\S$ 1.6.10  of \cite{nesterenko1}). The plots in Fig.~\ref{2SV_force} show the force $F_{in}$ at the contact between the striker and the first bead, the force $F(i)$ that denotes the mean of the contact force between particles $i$ and $i-1$ and that between $i+1$ and $i$, and the force $F_{out}$ recorded at the wall.  The observation time is 750 $\mu$s.  All of the beads are made of stainless steel (see the material properties in Table~\ref{tab:Mat1}). The $F(i)$ plots for $i>21$ in Fig.~\ref{2SV_force} clearly illustrate the aforementioned pulse disintegration phenomenon.  One can also see that the fitness $f = \| F_{out} \|_{L^{\infty}}$ of the 2SV is equal to 0.18 kN.

\begin{figure}[tbp]
    \centerline{\includegraphics[angle=0,width=85mm]{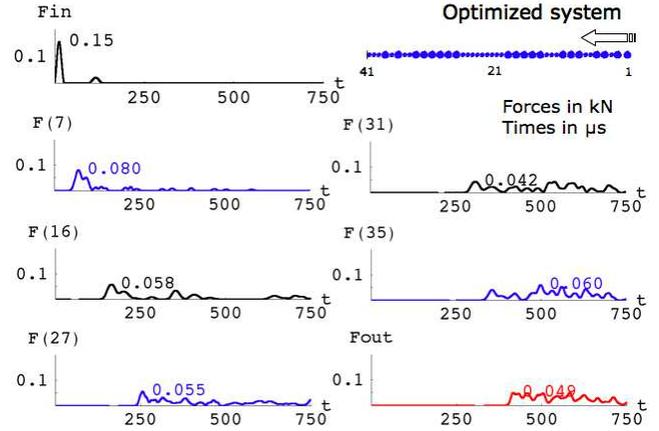}}
    \caption{\scriptsize{(Color online) Force versus time plots in the topology-optimized system.  (Compare this to the (unoptimized) stepped 2SV configuration in Fig.~\ref{2SV_force}.)}}\label{opt2SV_force}
\end{figure}

We ran a topology optimization of the 2SV by introducing $M=N=40$ genes $x_i$ related to the radius size (large or small) of the different beads.  (This does not include the striker--particle number 1--which is  assumed to have a large radius.)  We defined the genes so that $x_i \in [0,0.5]$ implies $r_{i+1}=r_S$, whereas $x_i \in (0.5,1]$ implies $r_{i+1}=r_L$. We used a penalty technique to constrain the number of particles with large and small radii to each be equal to 20; that is, we assigned a very large fitness $f$ to (unfeasible) solutions that do not satisfy this criterion.  We show the BGA-optimized system and the corresponding force-time plots in Fig.~\ref{opt2SV_force}. The optimized system has many large beads near the end of the chain that is hit by the striker (shown on the right), small beads near the wall, and an alternation of sequences of multiple consecutive large and small particles in the center of the system. We obtained a stable solution (i.e., a solution with constant best fitness) after about 340 generations of the algorithm.  Observe that pulse disintegration appears very early (within the first few beads) in the optimized system, so that the leading solitary wave transforms into a train of interacting, small-amplitude pulses.  This configuration exhibits a fitness of about 0.049 kN, which is almost four times smaller than that of the 2SV. 

\begin{figure}[htbp]
  \begin{center}
    \setlength{\unitlength}{1mm}
    $\begin{array}{cc}
    \epsfig{file=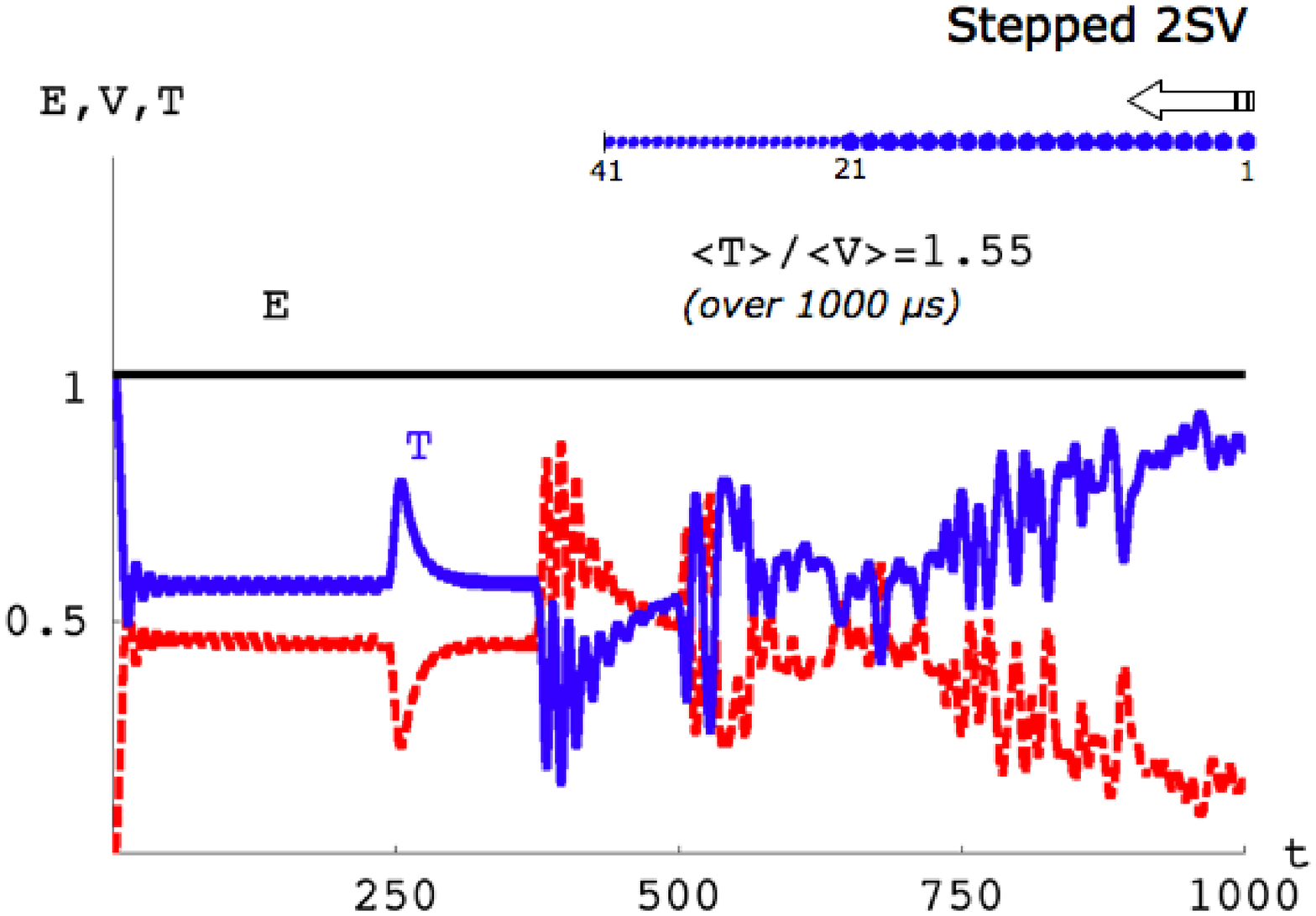,angle=0,width=44mm} &
    \epsfig{file=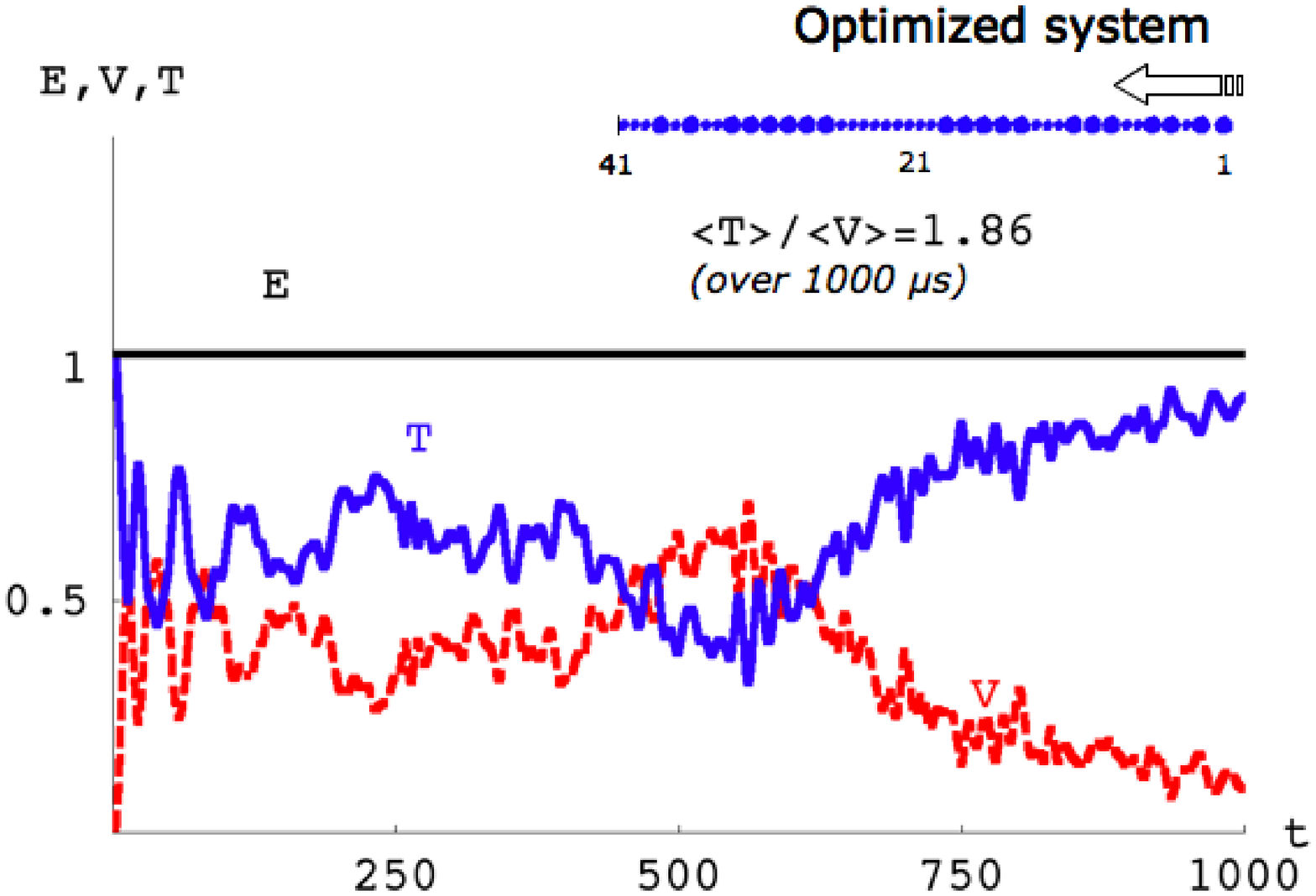,angle=0,width=44mm}       \end{array}$
     \caption{\scriptsize{(Color online) Energy versus time plots in the stepped 2SV and in the topology-optimized system (energies are in mJ, and times are in $\mu$ s).}}
     \label{2SV_energy}
  \end{center}
\end{figure}

We compare the energies (as a function of time) of the stepped 2SV and the optimized system in Fig.~\ref{2SV_energy} over a time window preceding the achievement of a loose state (in which there are no interaction forces).\footnote{In the absence of gravity and precompression, a loose state is reached after a sufficiently long time because the granular chain is constrained only at one end.  The dynamics evolve so that the interactions go to zero.}  We obtained this by restricting the energy time-histories up to the first instant $t>0$ for which $T = 0.99 {E}$. The kinetic energy $T$ of the 2SV shows a marked peak when the leading wave passes from the larger sub-chain to the smaller one and valleys when the wave is reflected at the wall.  The potential energy $V$ behaves in the opposite manner because the total energy is conserved.  In the optimized system, on the other hand, the valleys and peaks of $T$ and $V$ arise earlier during wave propagation, and the peaks of the potential energy remain markedly lower than those observed in the 2SV.  Denoting by $\langle T\rangle$ and $\langle V\rangle$ the time-averaged values of $T$ and $V$, respectively, over a time of 1000 $ms$ from the striker impact, we observe that in both the 2SV (which has $\langle T\rangle/\langle V\rangle \approx 1.55$) and the optimized system (which has $\langle T\rangle/\langle V\rangle \approx 1.86$), the ratio $\langle T\rangle/\langle V\rangle$ deviates from the value 1.25 predicted by the virial theorem of statistical physics \cite{pathria}.  Nesterenko observed similar results using randomized granular chains subject to piston-like impacts \cite{nesterenko1}.

\begin{figure}[htbp]
  \begin{center}
    \setlength{\unitlength}{1mm}
    $\begin{array}{cc}
    \epsfig{file=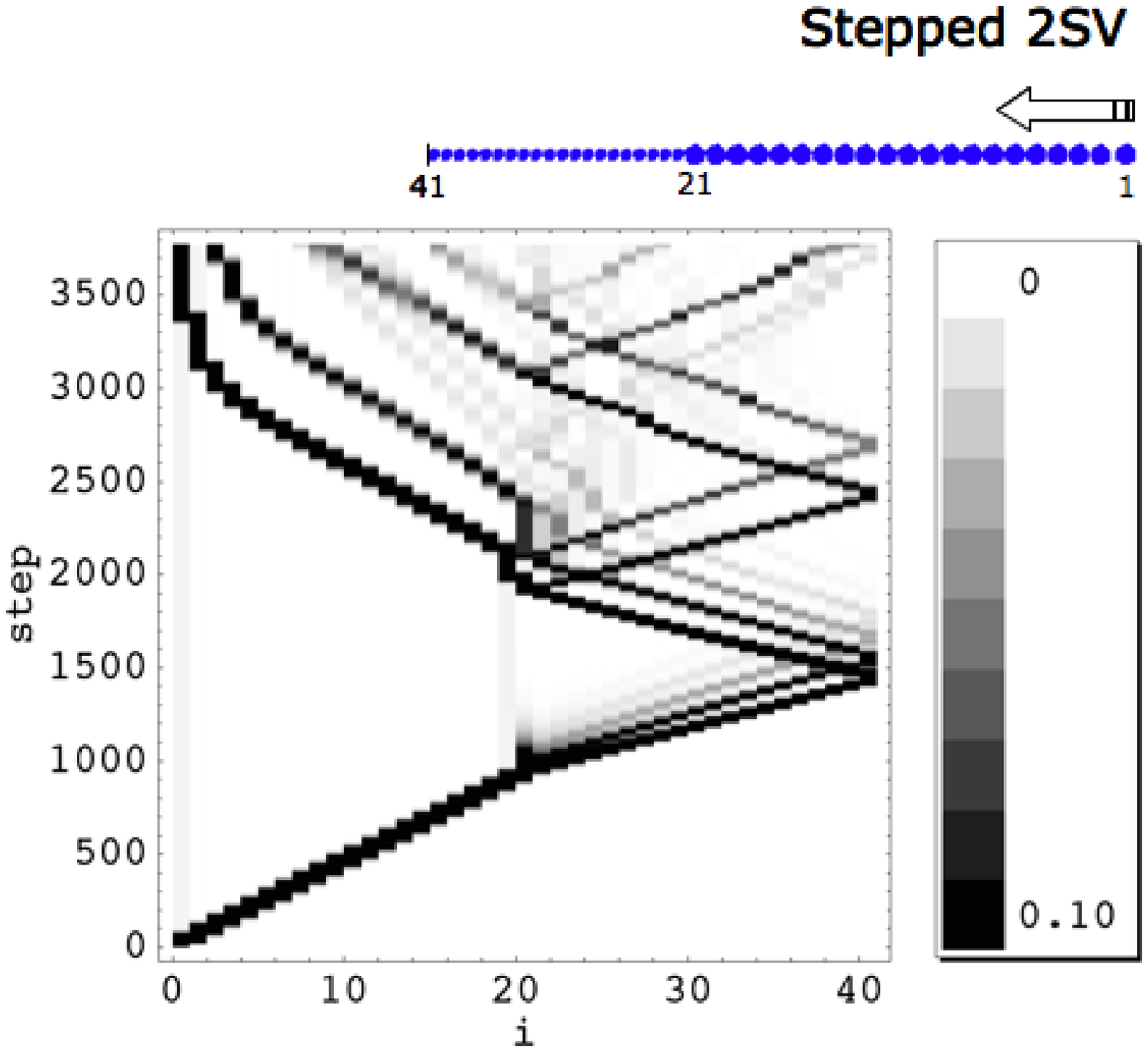,angle=0,width=44mm} &
    \epsfig{file=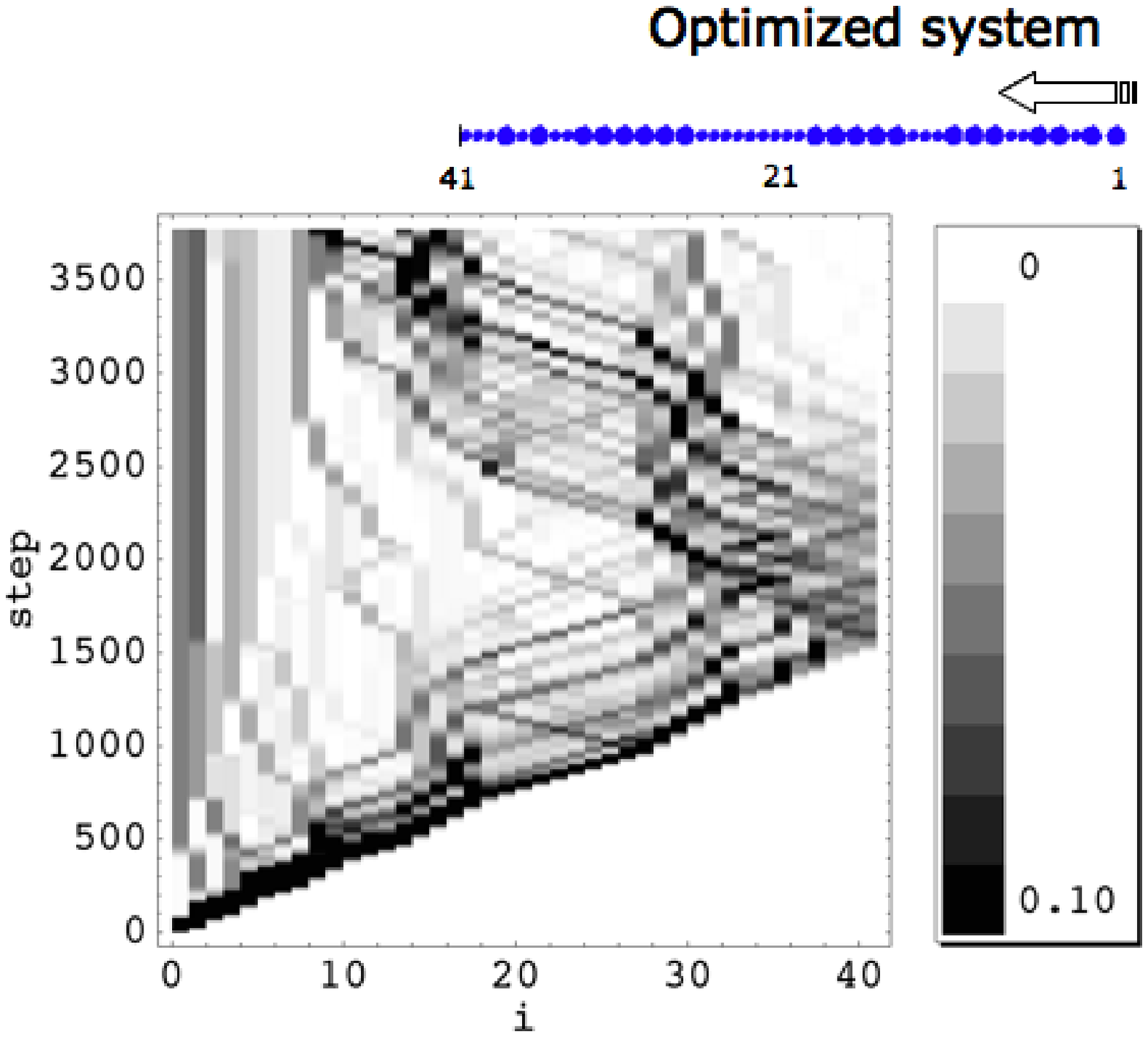,angle=0,width=44mm}       \end{array}$
     \caption{\scriptsize{(Color online) Density plots of particle energies normalized to unity.  Horizontal axes are labeled according to particle site and vertical axes give the time step.}}
     \label{2SV_density}
  \end{center}
\end{figure}

\begin{figure}[tbp]
    \centerline{\includegraphics[angle=0,width=60mm]{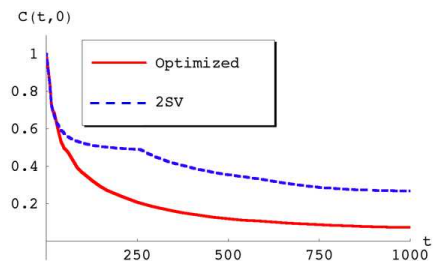}}
    \caption{\scriptsize{(Color online) Energy correlation function versus time for the topology-optimized and stepped 2SV systems.}}\label{topology_corr}
\end{figure}

Figure \ref{2SV_density} shows density plots of particle energies $E_i$ for the stepped 2SV and the optimized system. In each plot, the horizontal axis indicates the particle site, the vertical axis shows the progressive step number (we produced a plot for every five integration steps), and the shading gives a density plot of the energy $E_i$ normalized to unity (i.e., the energy divided by its maximum value, among all the beads). One can clearly recognize the impulse disintegration phenomenon in the stepped 2SV when the incident pulse passes from the large-bead regime to the small-bead regime.  In this system, pulse reflection occurs only at the wall (and not along the body of the chain) during the first transmission.  Note that when already-reflected pulses pass from the small-bead regime to the large-bead regime, they are reflected for a second time.  In the optimized system, one observes a combination of disintegration and reflection of traveling pulses along the entire chain.  One also observes the production of interacting pulses that travel in opposite directions. In particular, the pulses moving toward the wall combine to form an extended (long-wavelength), small-amplitude wave that is clearly visible in Fig.~\ref{opt2SV_force}. 
Figure \ref{topology_corr} shows the time histories of the energy correlation function for the stepped 2SV and topology-optimized systems, revealing that the latter exhibits a faster and stronger thermalization (i.e., equipartition of energy) than the former.

\nobreak

\nobreak

\nobreak

\subsection{Size optimization}\label{size}

\begin{figure}[tbp]
    \centerline{\includegraphics[angle=0,width=80mm]{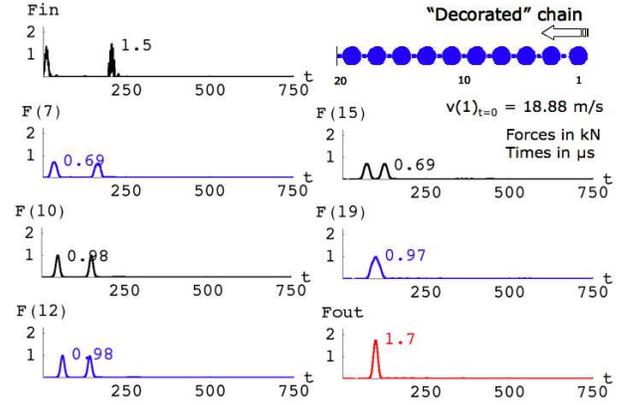}}
    \caption{\scriptsize{(Color online) Force versus time plots in a decorated chain.}}\label{dec20_force}
\end{figure}

\begin{figure}[tbp]
    \centerline{\includegraphics[angle=0,width=80mm]{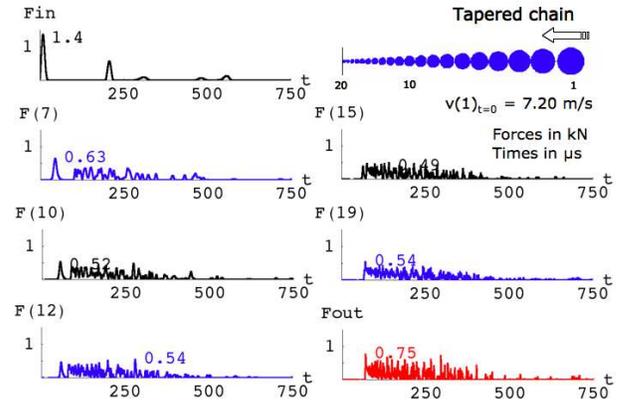}}
    \caption{\scriptsize{(Color online) Force versus time plots in a tapered chain.}}\label{tap20_force}
\end{figure}

Doney and Sen recently studied the energy absorption capabilities of 1D granular protectors consisting of ``tapered" and/or ``decorated" chains \cite{donsen05,donsen06}.  The simplest type of tapered chain is composed of a sequence of progressively larger or progressively smaller beads, and a decorated chain is a composite chain obtained by placing interstitial small grains between the large grains in a monodisperse chain.  Using analytical and numerical techniques, Doney and Sen observed marked energy absorption in highly tapered chains.  For their analytical work, they employed the hard-sphere approximation.  In their numerical studies, they utilized hydrocode simulations up to very high impact velocities (up to 1 km/s).  In Figs.~\ref{dec20_force} and \ref{tap20_force}, we show several numerical force recordings that correspond, respectively, to a decorated and a tapered chain impacted by strikers with different velocities.  In both cases, the initial force peak is $F_{in} = 1.4$ kN. The decorated chain consists of a sequence of dimers formed by alternating $r=\hat{r}_L=3.243$ mm and $r=\hat{r}_S=0.973$ mm stainless steel beads with total length $7.78$ cm (not counting the striker).  The striker (3.243 mm) impacts this chain with an initial velocity of 18.88 m/s. 
The tapered chain, on the other hand, is composed of 20 stainless steel beads with decreasing radii, ranging from $r=r_L=5$ mm (the striker, which impacts the chain at a speed of 7.20 m/s) to $r=r_S=0.675$ mm over a length of $7.78$ cm (not counting the striker).  That is, the tapering ratio is $q_s = r_{i+1}/r_i = 0.1$.  
 The fitness of the decorated chain is about 1.7 kN, whereas that of the tapered chain is equal to about 0.75 kN.

\begin{figure}[tbp]
    \centerline{\includegraphics[angle=0,width=80mm]{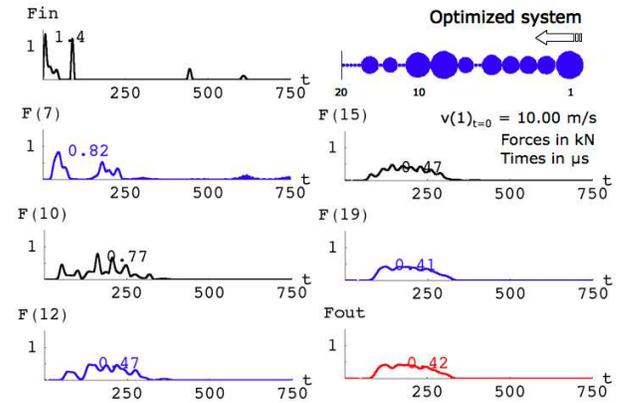}}
    \caption{\scriptsize{(Color online) Force versus time plots in the size-optimized system.}}\label{opt20_force}
\end{figure}

\begin{figure}[htbp]
  \begin{center}
    \setlength{\unitlength}{1mm}
    $\begin{array}{cc}
    \multicolumn{2}{c}{\epsfig{file=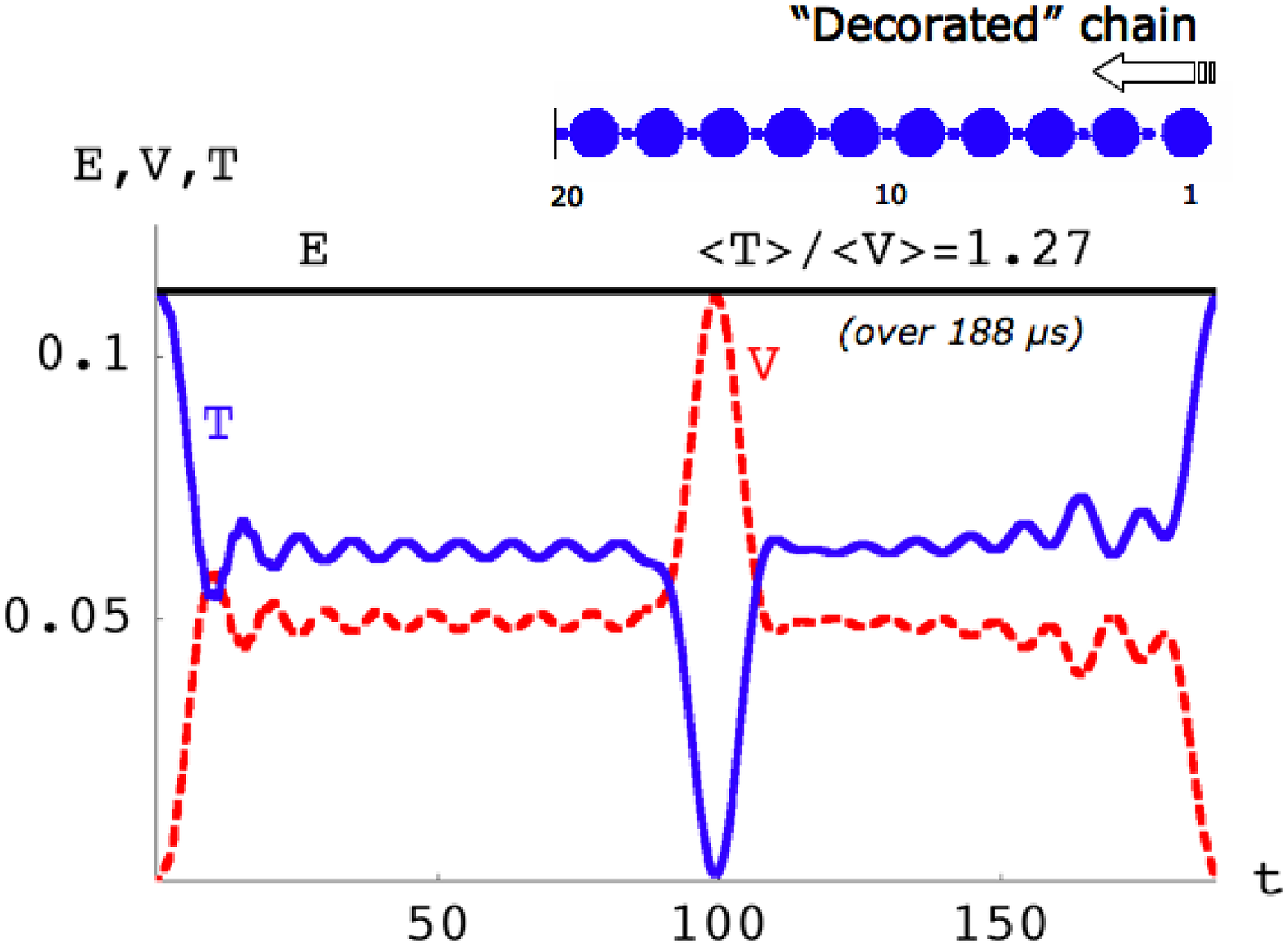,angle=0,width=44mm}}\\
    \epsfig{file=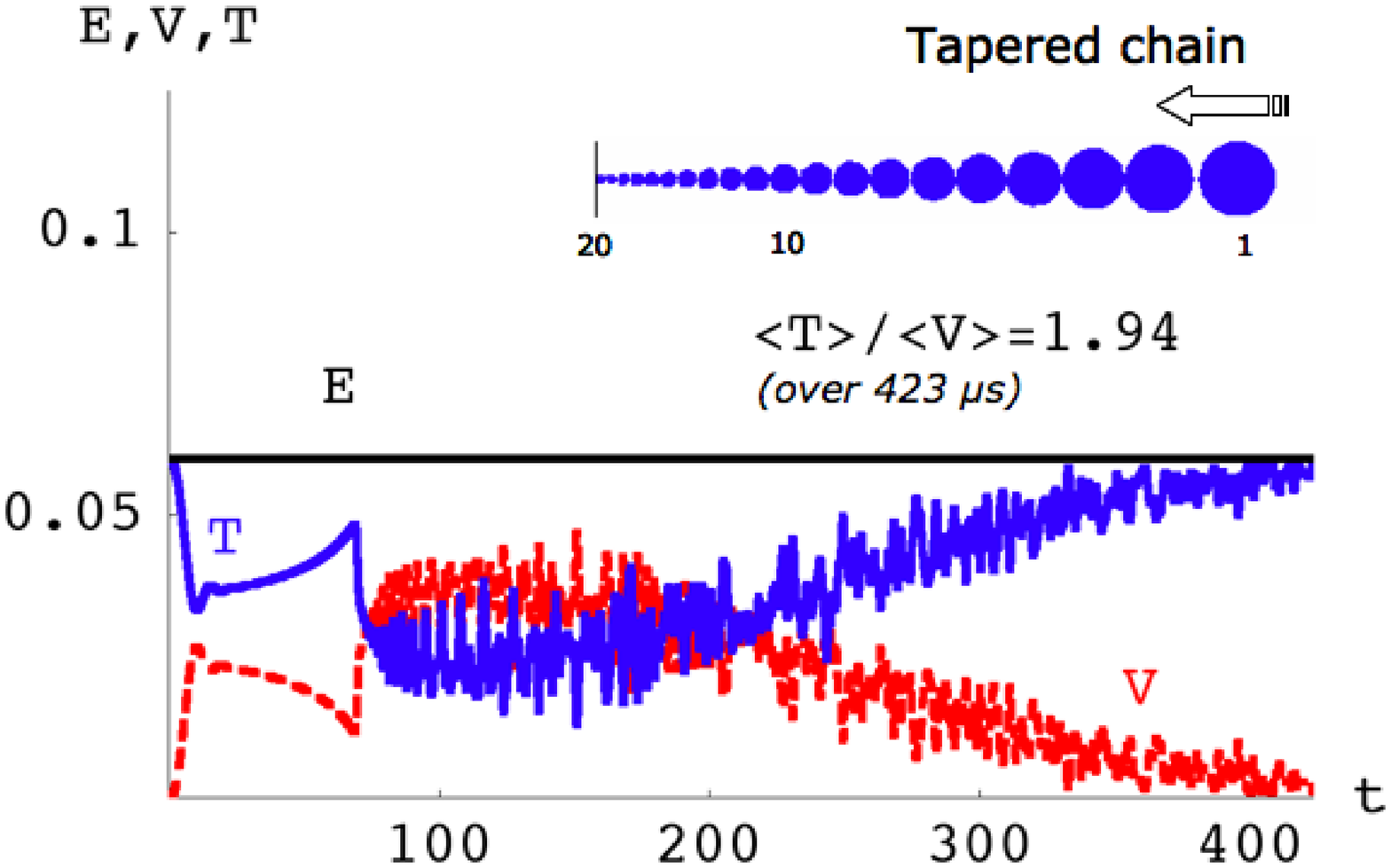,angle=0,width=44mm} &
    \epsfig{file=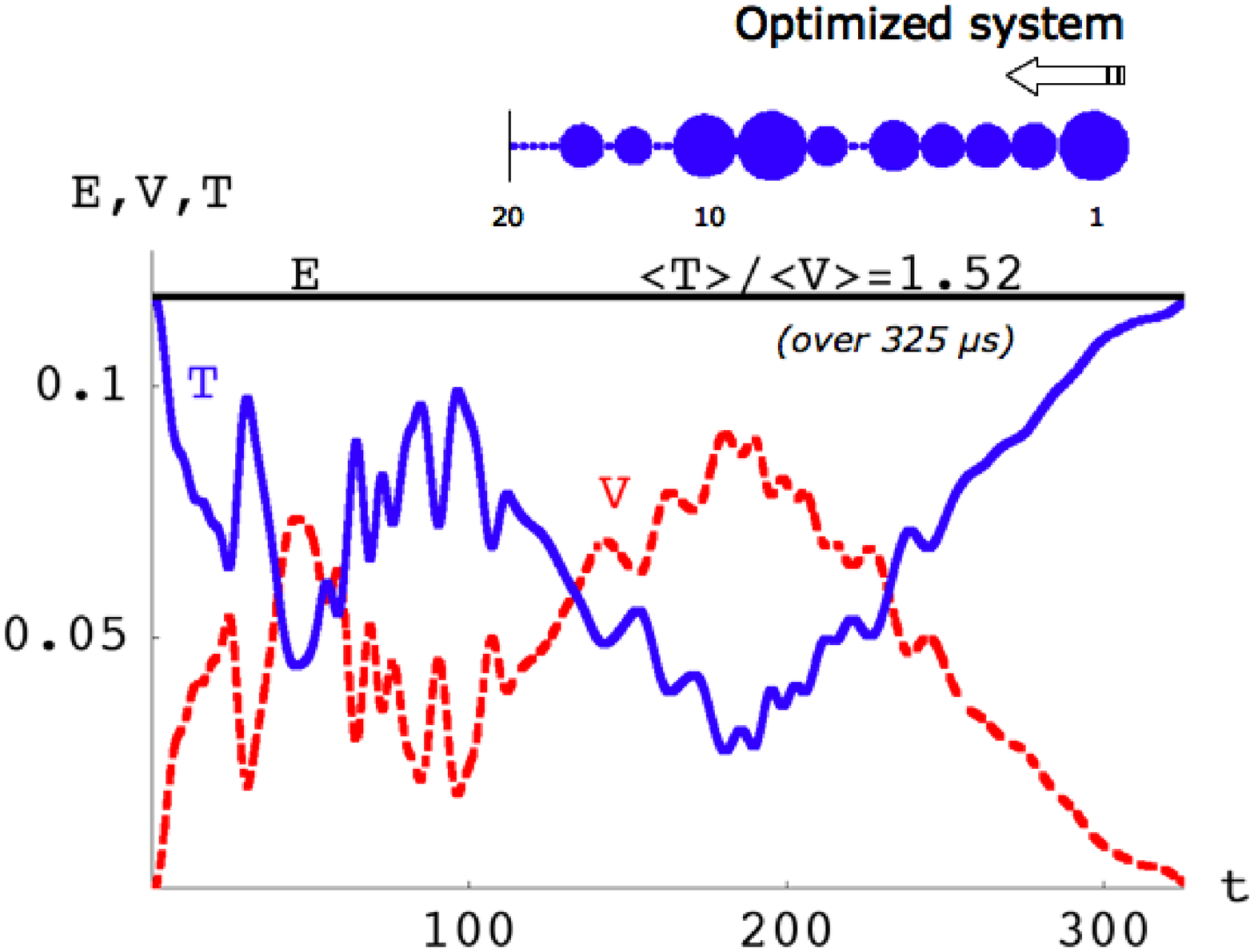,angle=0,width=44mm}       \end{array}$
     \caption{\scriptsize{(Color online) Energy versus time plots in the decorated, tapered, and size-optimized chains (energies in J, times in $\mu$ s).}}
     \label{size_energy}
  \end{center}
\end{figure}

\begin{figure}[htbp]
  \begin{center}
    \setlength{\unitlength}{1mm}
    $\begin{array}{cc}
    \multicolumn{2}{c}{\epsfig{file=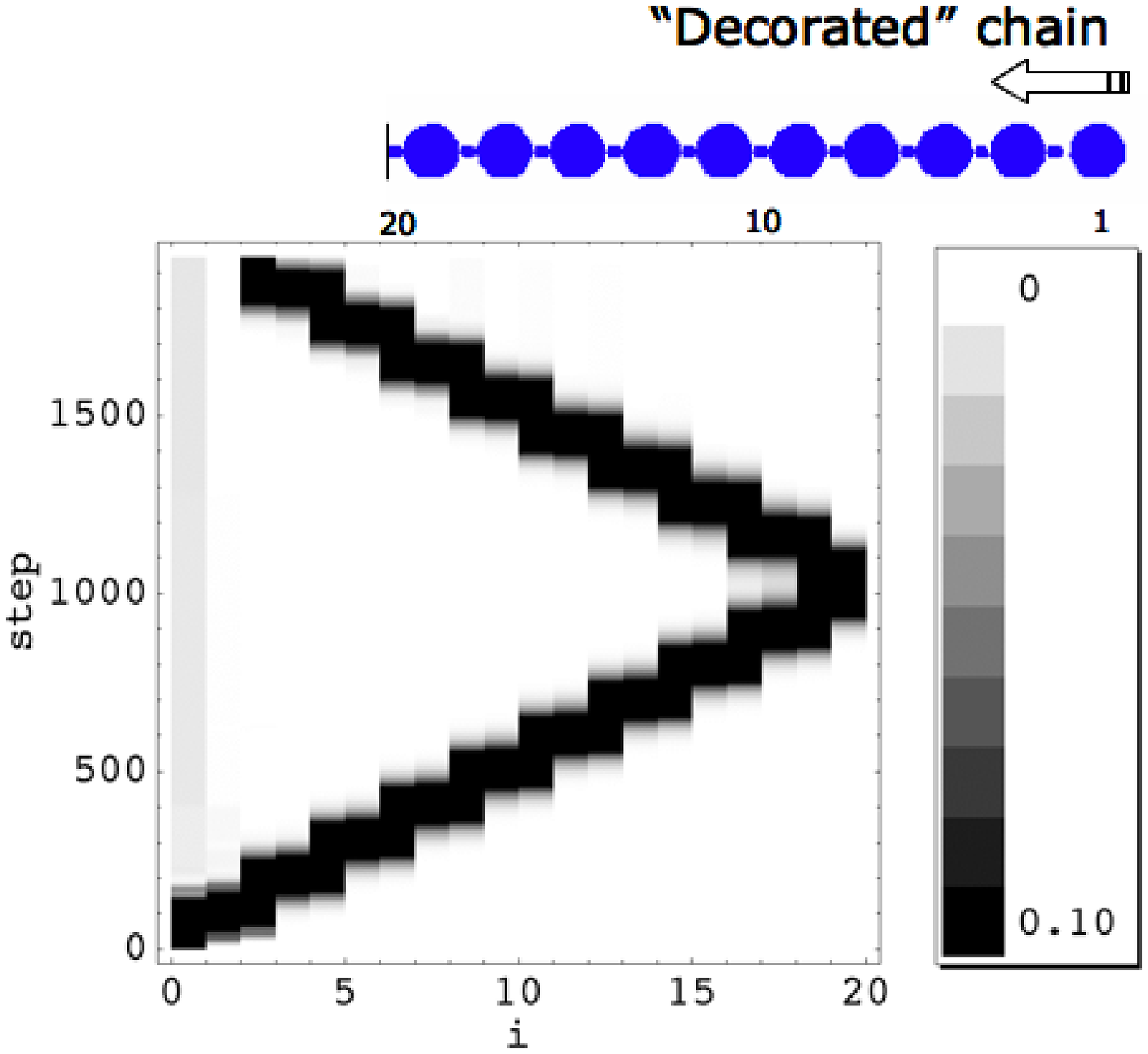,angle=0,width=44mm}}\\
    \epsfig{file=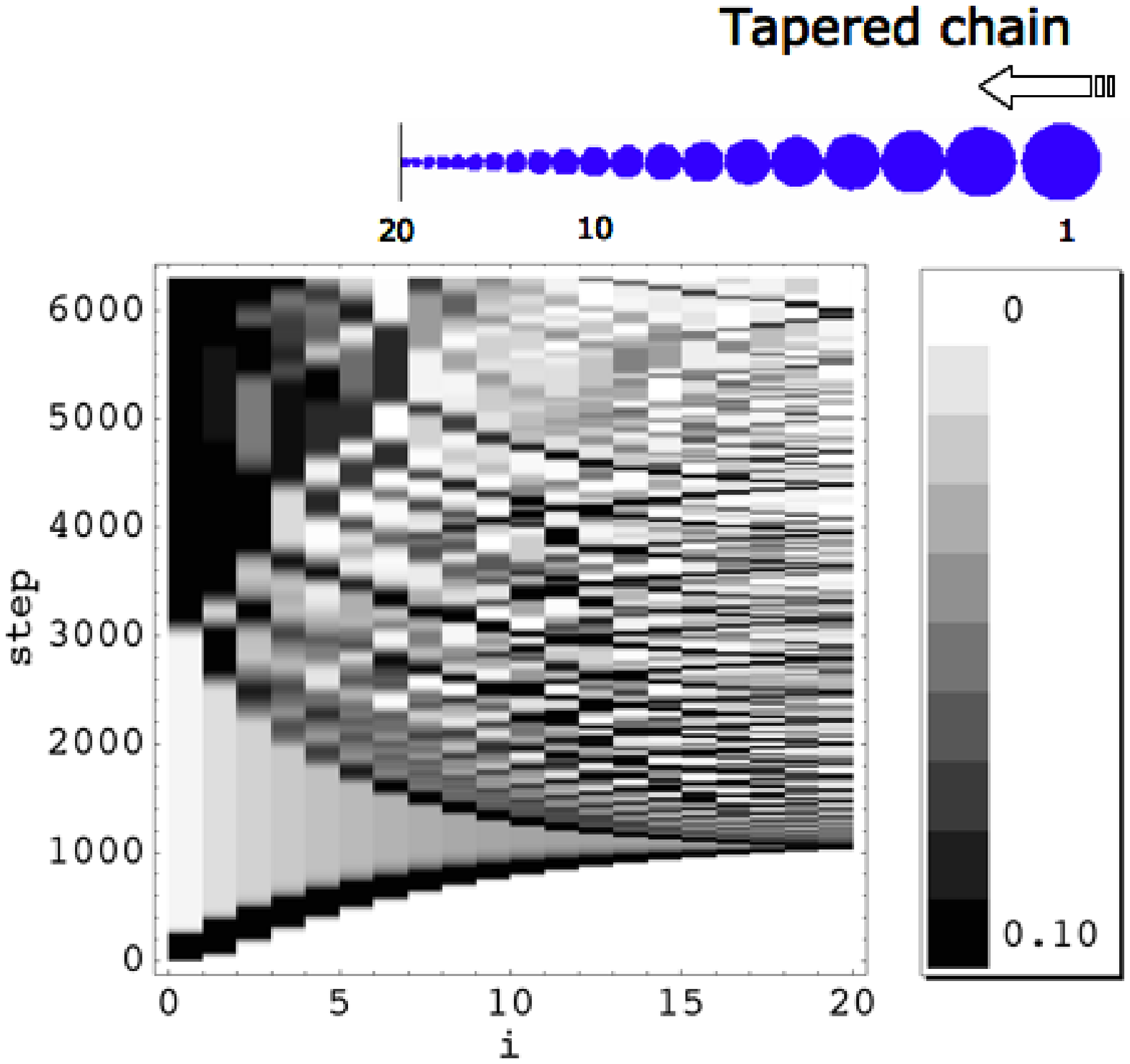,angle=0,width=44mm} &
    \epsfig{file=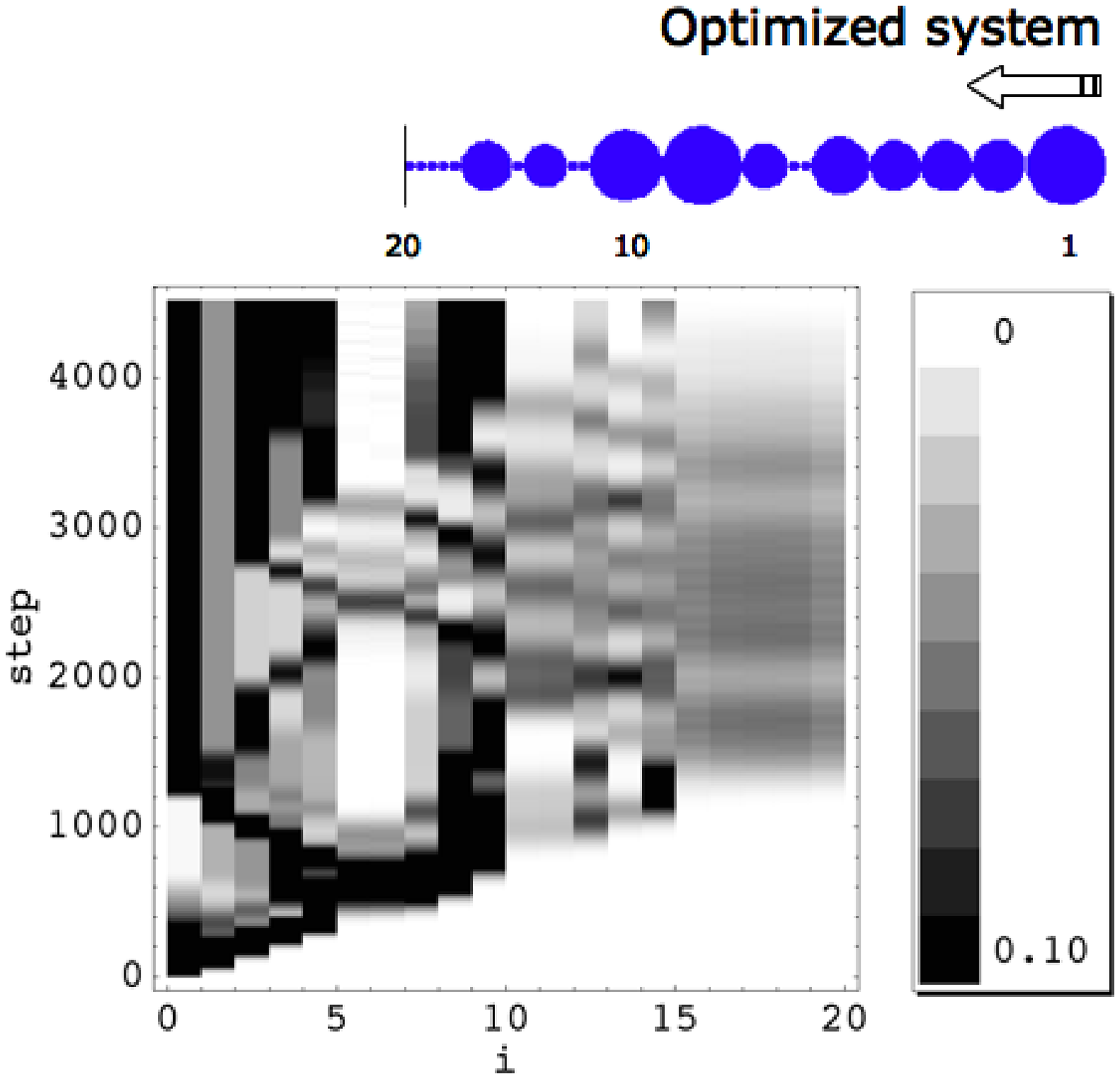,angle=0,width=44mm}       \end{array}$
     \caption{\scriptsize{(Color online) Density plots of particle energies normalized to unity.  Horizontal axes are labeled according to particle site and vertical axes give the time step.}}
     \label{size_density}
  \end{center}
\end{figure}

\begin{figure}[tbp]
    \centerline{\includegraphics[angle=0,width=60mm]{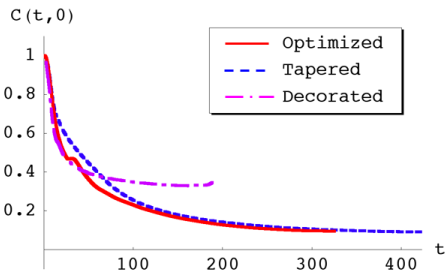}}
    \caption{\scriptsize{(Color online) Energy correlation function versus time for the examined systems.}}\label{size_corr}
\end{figure}

We carried out a size optimization of the above systems, introducing $M=19$ genes $x_i$ related to the radii of the different beads (not including the striker), which in the present example (in contrast to the previous one) were assumed to change continuously in the interval $[r_S, r_L]$ (with $r_{i+1} = r_S + x_i (r_L - r_S)$). We ran a BGA optimization with a striker radius always equal to 5 mm and an impact speed equal to 10 m/s.  We also constrained the total length of the system to remain equal to 7.88 cm (not including the striker). We obtained constant best fitness and the solution shown in Fig.~\ref{opt20_force} after about 590 generations. The fitness of the size-optimized system (0.42 kN) is about 1.8 times smaller than that of the (unoptimized) tapered chain. Observe that there is simple reflection at the wall in the decorated chain (see Fig.~\ref{dec20_force}), significant pulse disintegration in the tapered chain (see Fig.~\ref{tap20_force}), and a transformation of the leading solitary pulses into an extended (long-wavelength), small-amplitude wave in the optimized chain (Fig.~\ref{opt20_force}).  The choice of the fitness parameters used in this study differs from Doney and Sen's \cite{donsen06}, which instead minimizes the kinetic energy ratio $K_{out}/K_{in}$ between output and input.  Because of the continuum formulation of the genes in the current examples, our work encompasses all geometries considered in Ref.~\cite{donsen06}.  We compare the energy profiles of the decorated, tapered, and optimized chains in Fig.~\ref{size_energy}. We observed the highest $\langle T\rangle/\langle V\rangle$ in the tapered chain ($\langle T\rangle/\langle V\rangle \approx 1.94$), resulting from the anticipated evolution of this system toward a loose state.  Figure \ref{size_density} shows the density plots of particle energies for the three systems under examination. As in the previous example, one can clearly observe from the plots the effects of combined wave disintegration and reflection in the optimized system. The profiles of the energy correlation function shown in Fig.~\ref{size_corr} indicate that the size-optimized and the tapered chains both evolve toward thermalization, with slightly faster decay of $C(t,0)$ in the former system. We show in the next section that a similar behavior can also be induced in a long dimeric system (i.e., in a long decorated chain) by introducing suitable alterations of the periodic particle arrangement (i.e., by introducing another form of disorder into the system ).

\nobreak

\nobreak

\nobreak

\nobreak

\subsection{Periodic sequences of optimized cells}\label{period}

\begin{figure}[htbp]
  \begin{center}
    \setlength{\unitlength}{1mm}
    $\begin{array}{cc}
    \epsfig{file=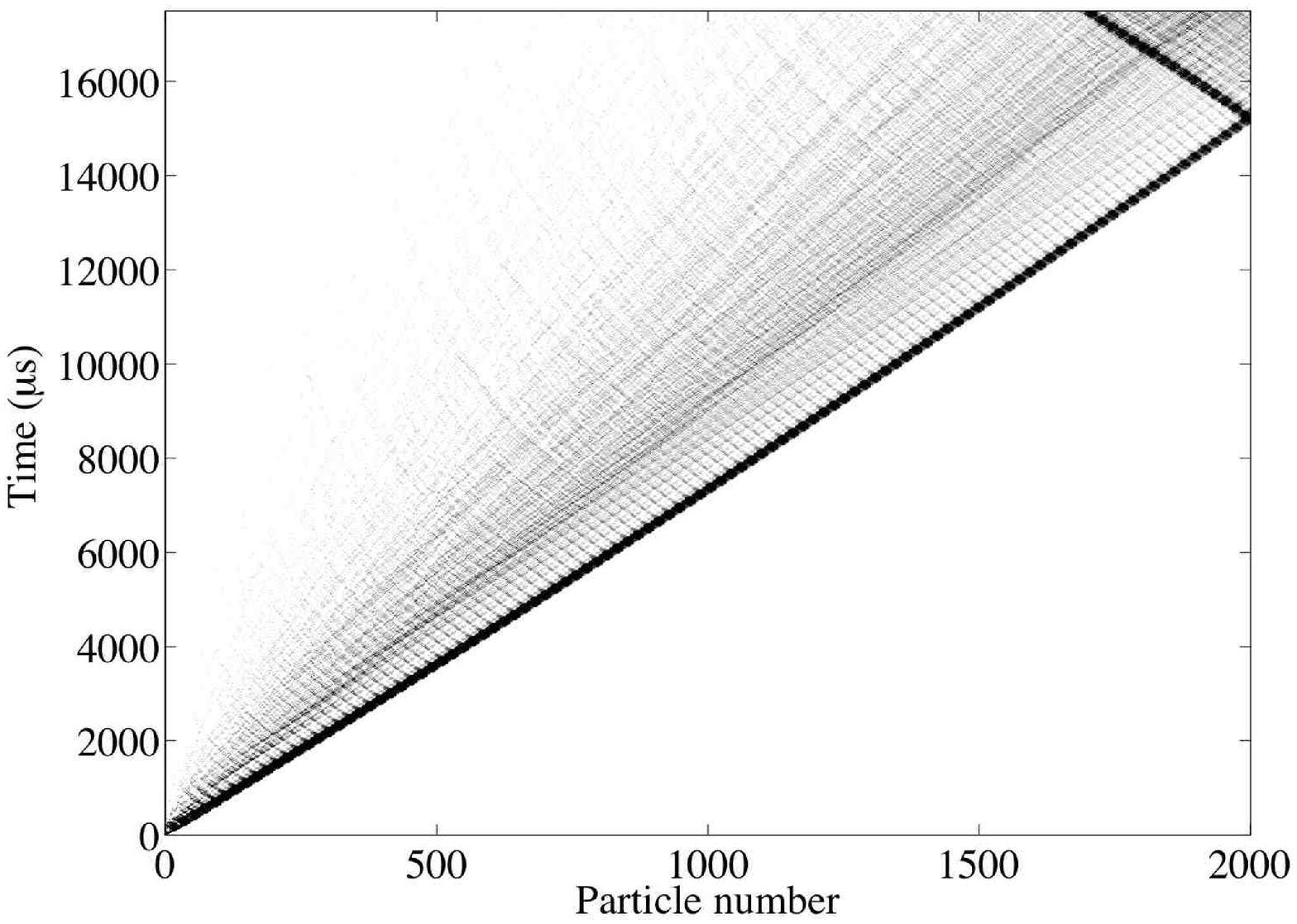,angle=0,width=44mm} &
    \epsfig{file=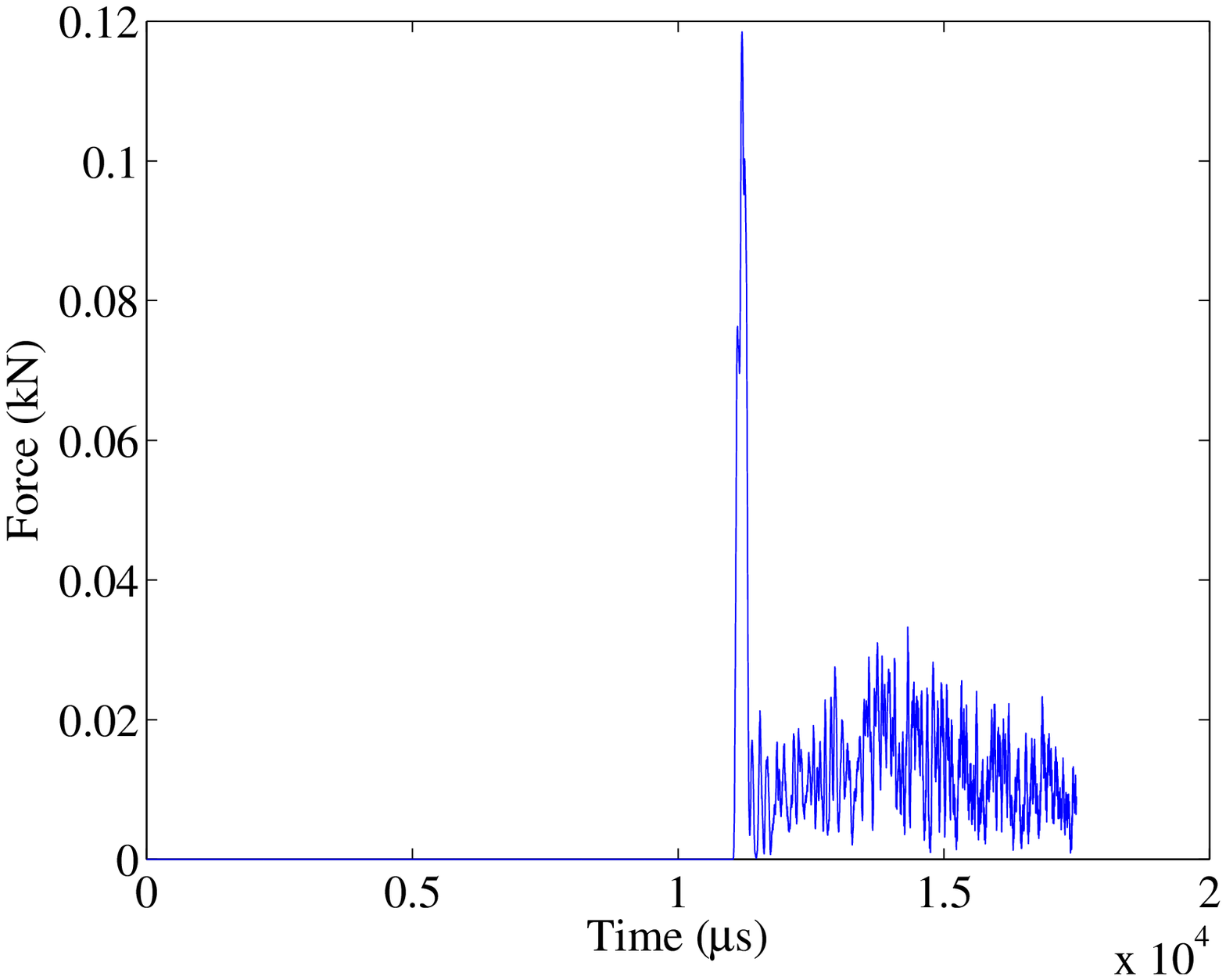,angle=0,width=44mm}       \end{array}$
     \caption{\scriptsize{(Left) Density plot (colored by force, with larger values given by darker shading) of the optimized chain in Fig.~\ref{opt20_force}. (Right) Force (in kN) versus time plot for particle 1500 of this chain.}}\label{optspacetime}
  \end{center}
\end{figure}

We now consider periodic sequences of the 19-particle size-optimized cells in the decorated chain shown in Fig.~\ref{opt20_force}.  As discussed above, the single optimized cells that we obtained can be viewed as disordered configurations, so it is interesting to investigate the effects on the wave dynamics of periodically repeating such a structure to obtain a ``quasi-disordered" configuration.  As can be seen in Fig.~\ref{optspacetime}, a reasonably localized wave structure does develop as long as there are enough periods, just as with periodic arrangements of simpler cells \cite{dimer,dimerlong}.  However, the original wave disintegrates and emits a significant number of secondary pulses, so that a stable coherent structure does not form.

\begin{figure}[htbp]
  \begin{center}
    \setlength{\unitlength}{1mm}
    $\begin{array}{cc}
    \epsfig{file=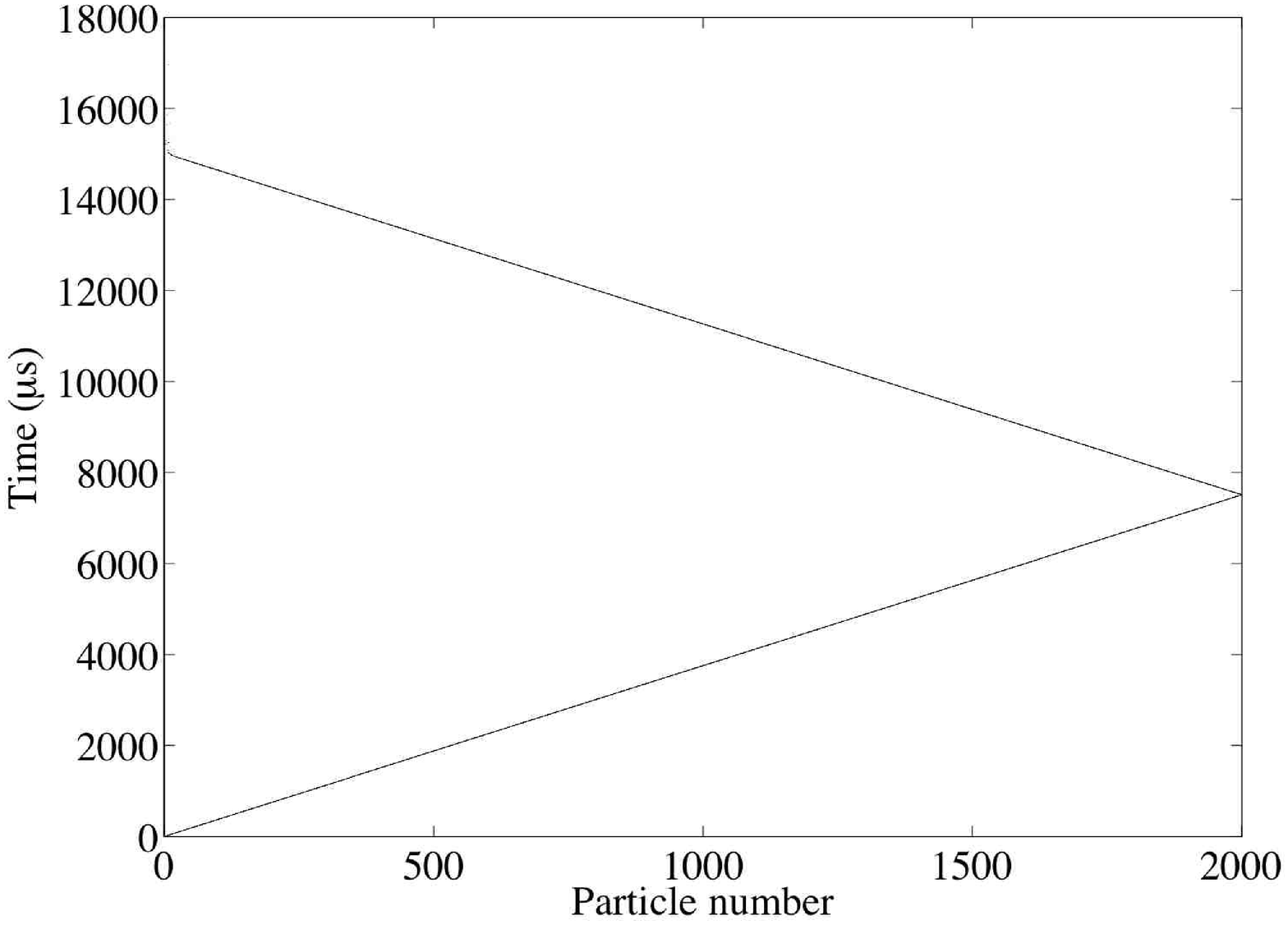,angle=0,width=44mm} &
    \epsfig{file=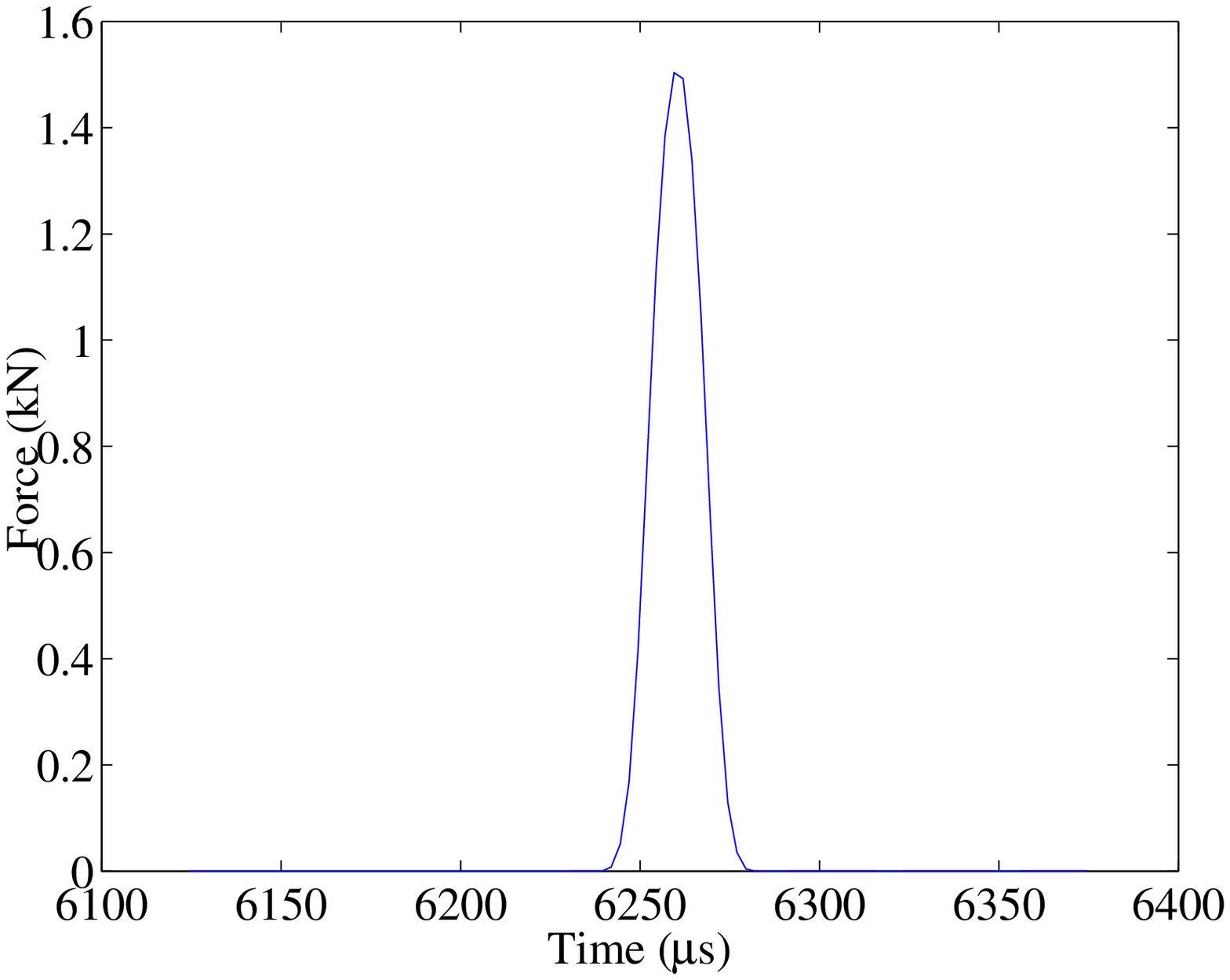,angle=0,width=44mm}       \end{array}$
  $\begin{array}{cc}
    \epsfig{file=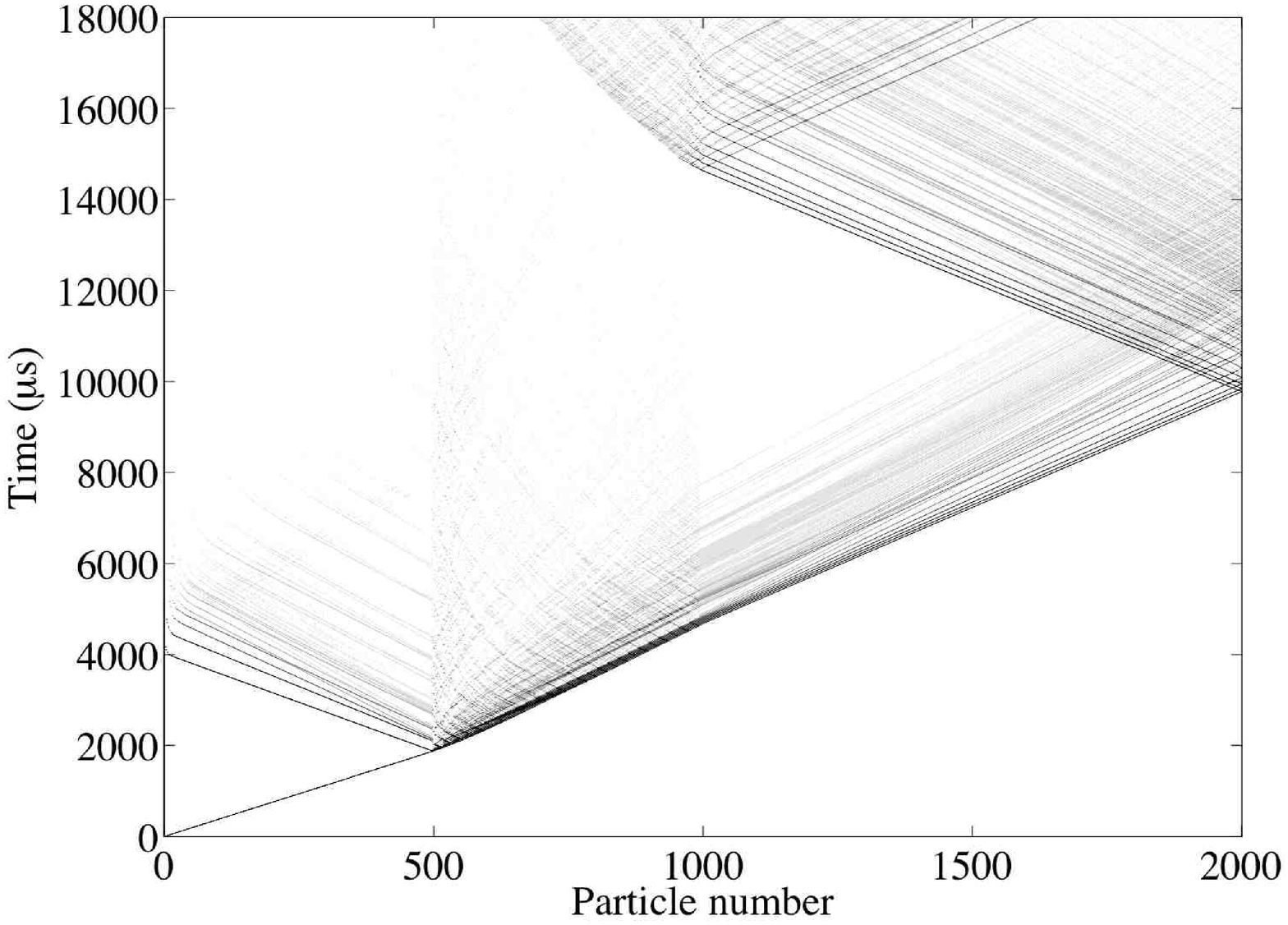,angle=0,width=44mm} &
    \epsfig{file=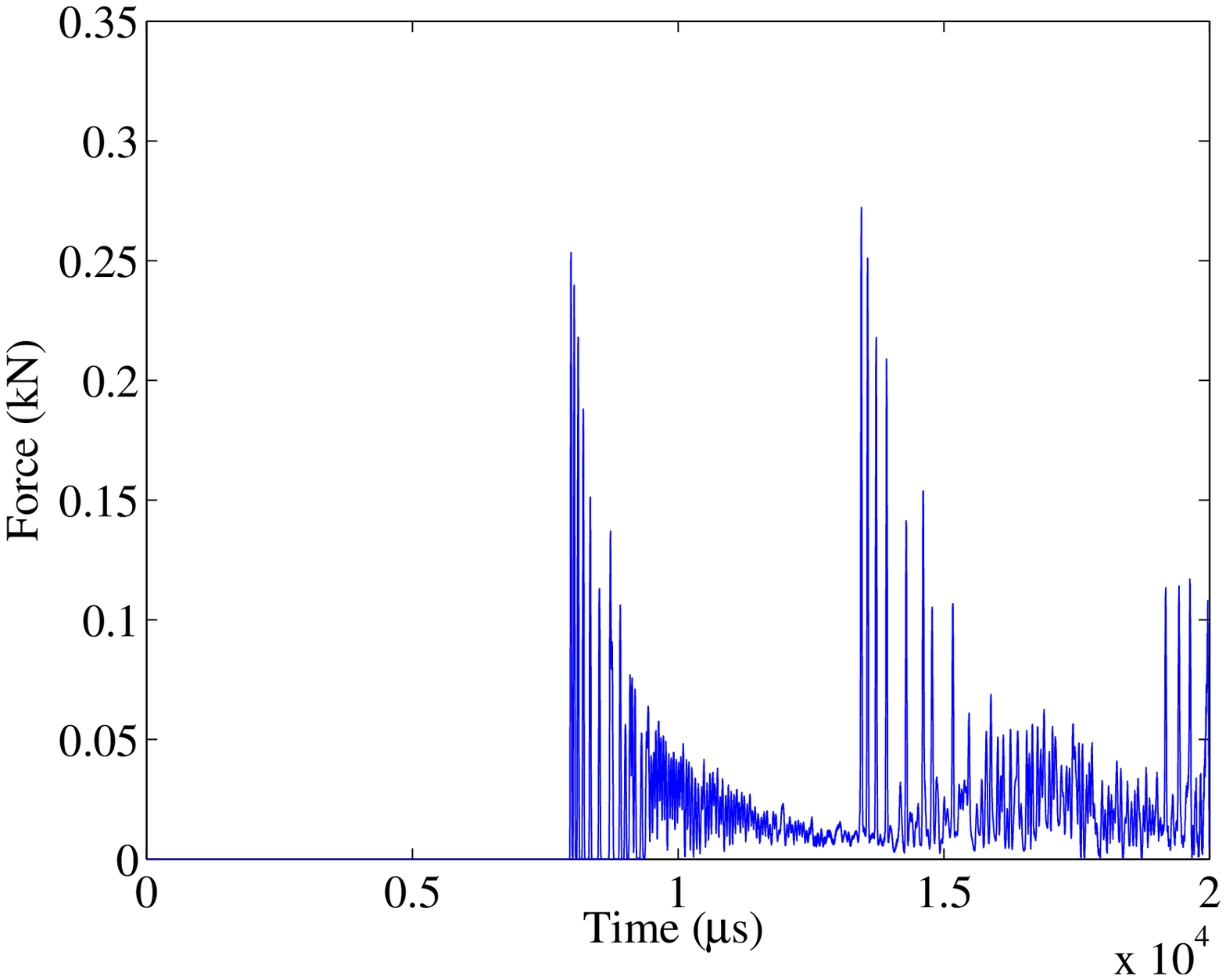,angle=0,width=44mm}       \end{array}$
  $\begin{array}{cc}
    \epsfig{file=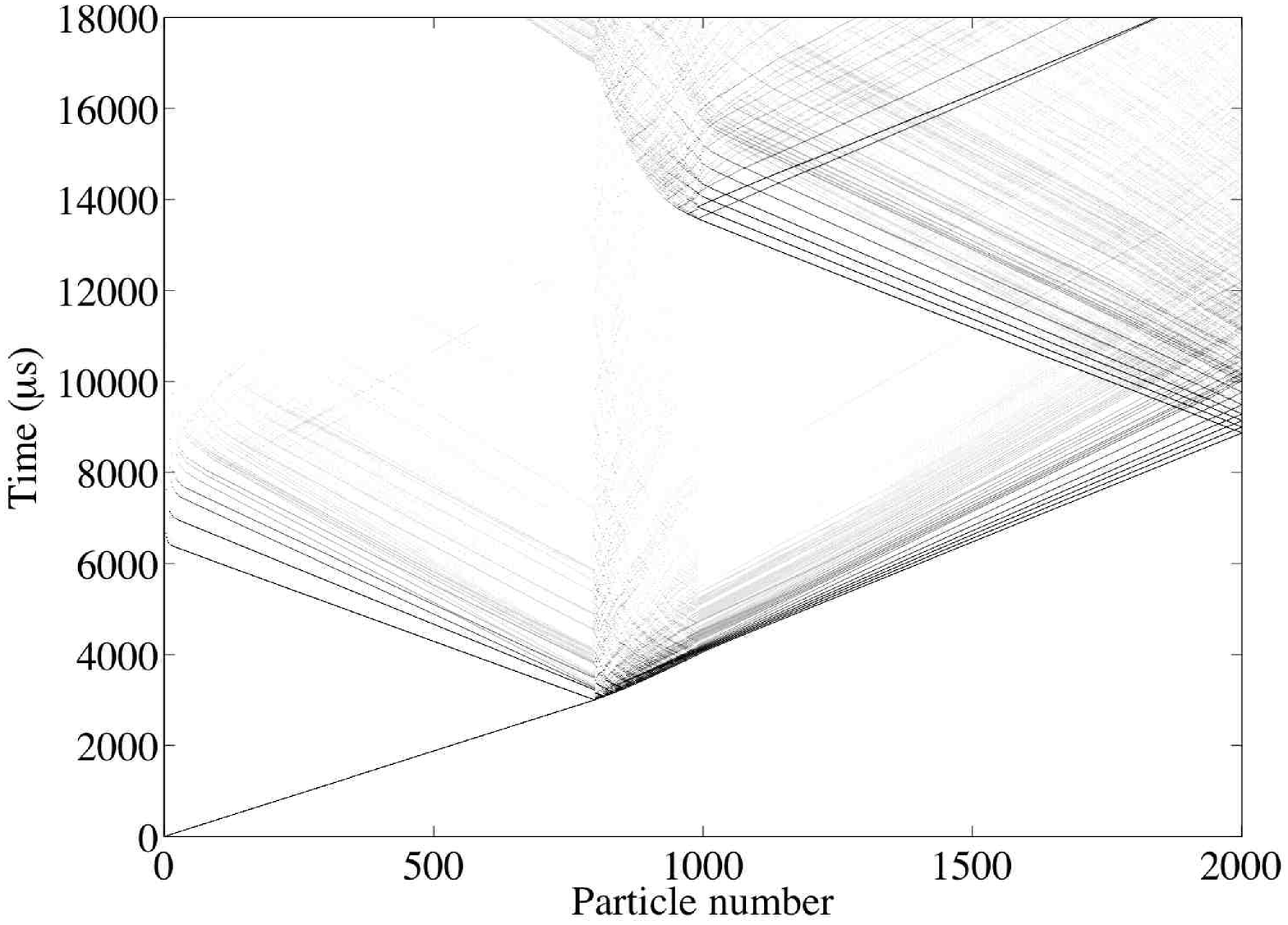,angle=0,width=44mm} &
    \epsfig{file=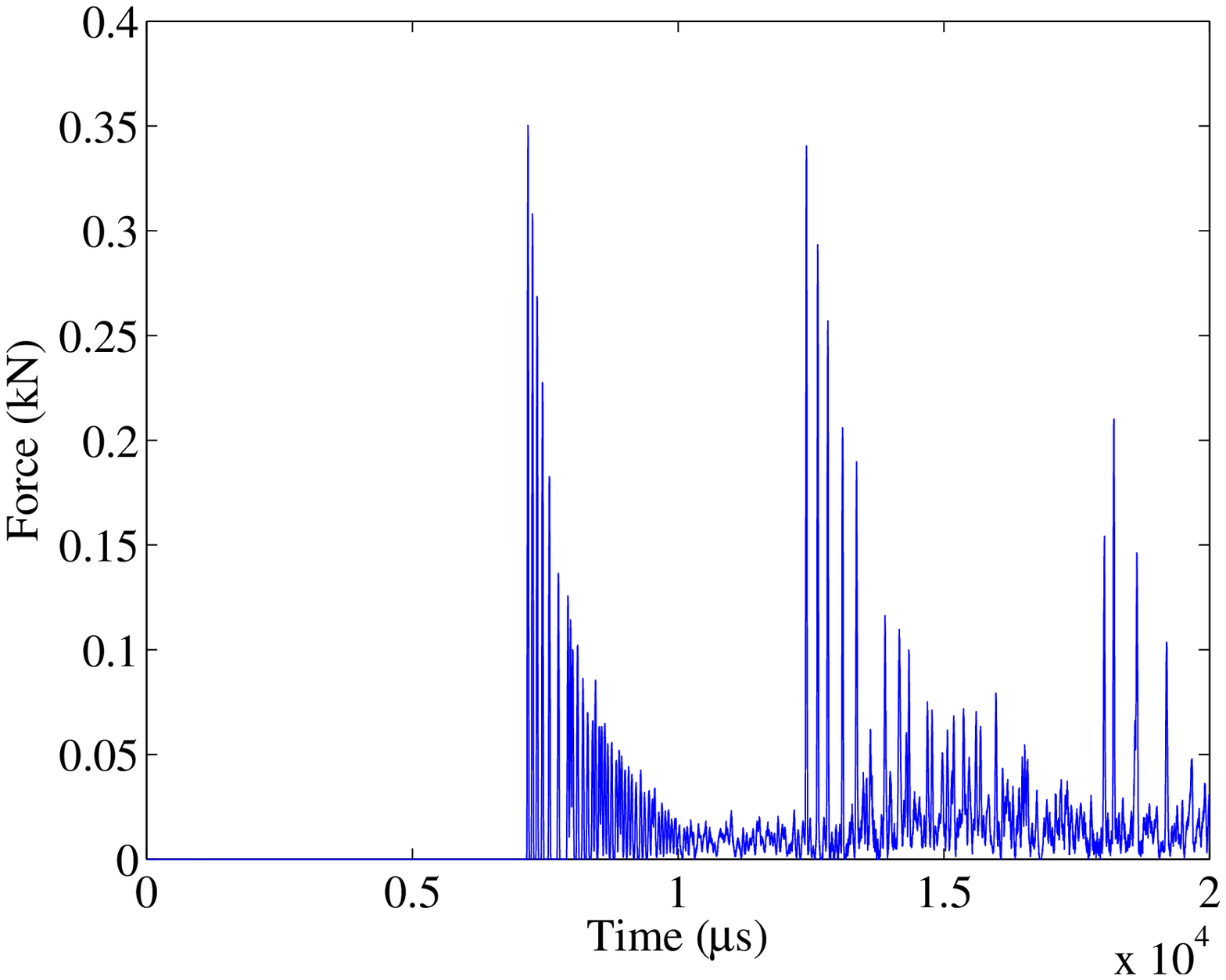,angle=0,width=44mm}       \end{array}$
   $\begin{array}{cc}
    \epsfig{file=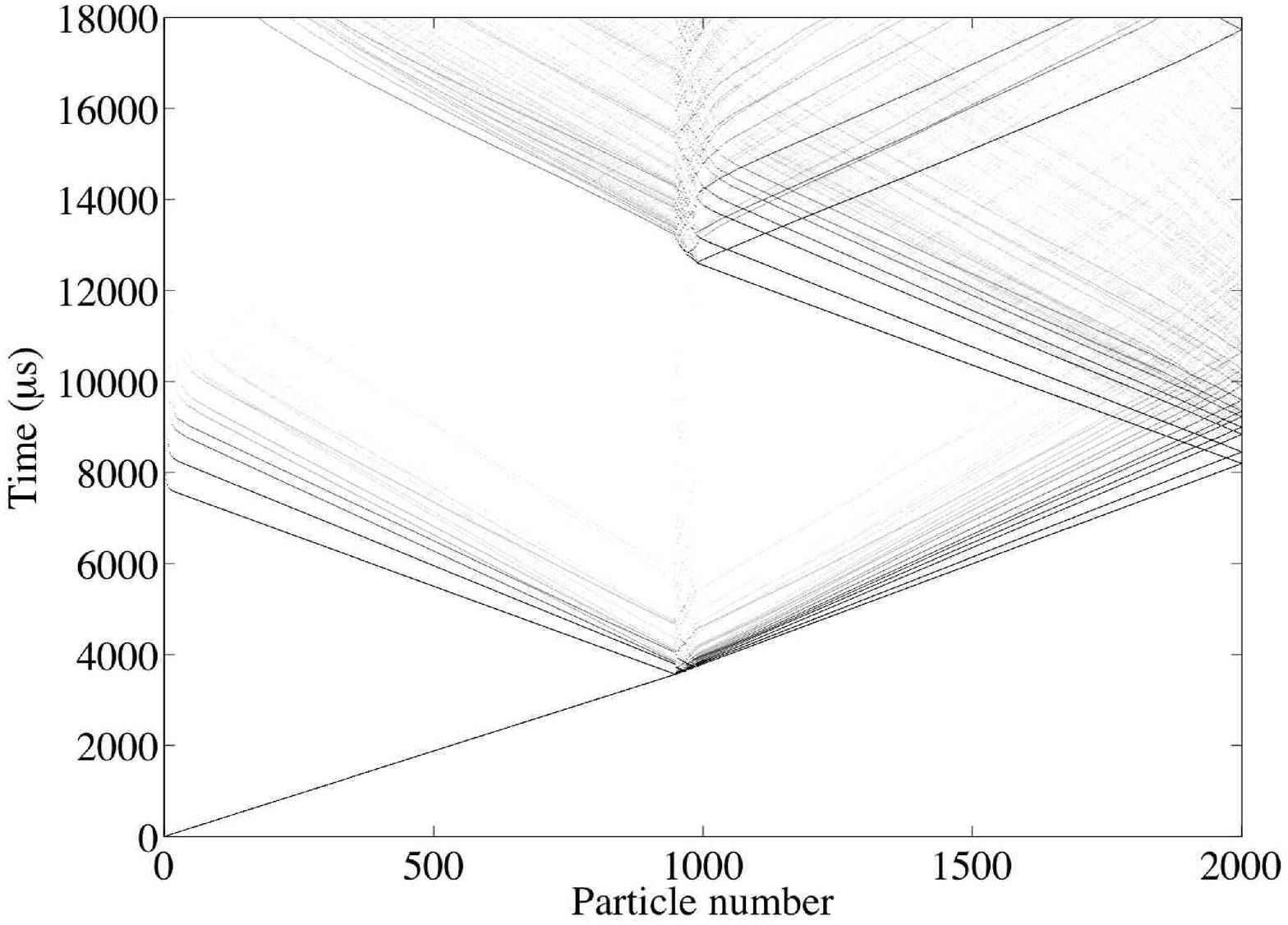,angle=0,width=44mm} &
    \epsfig{file=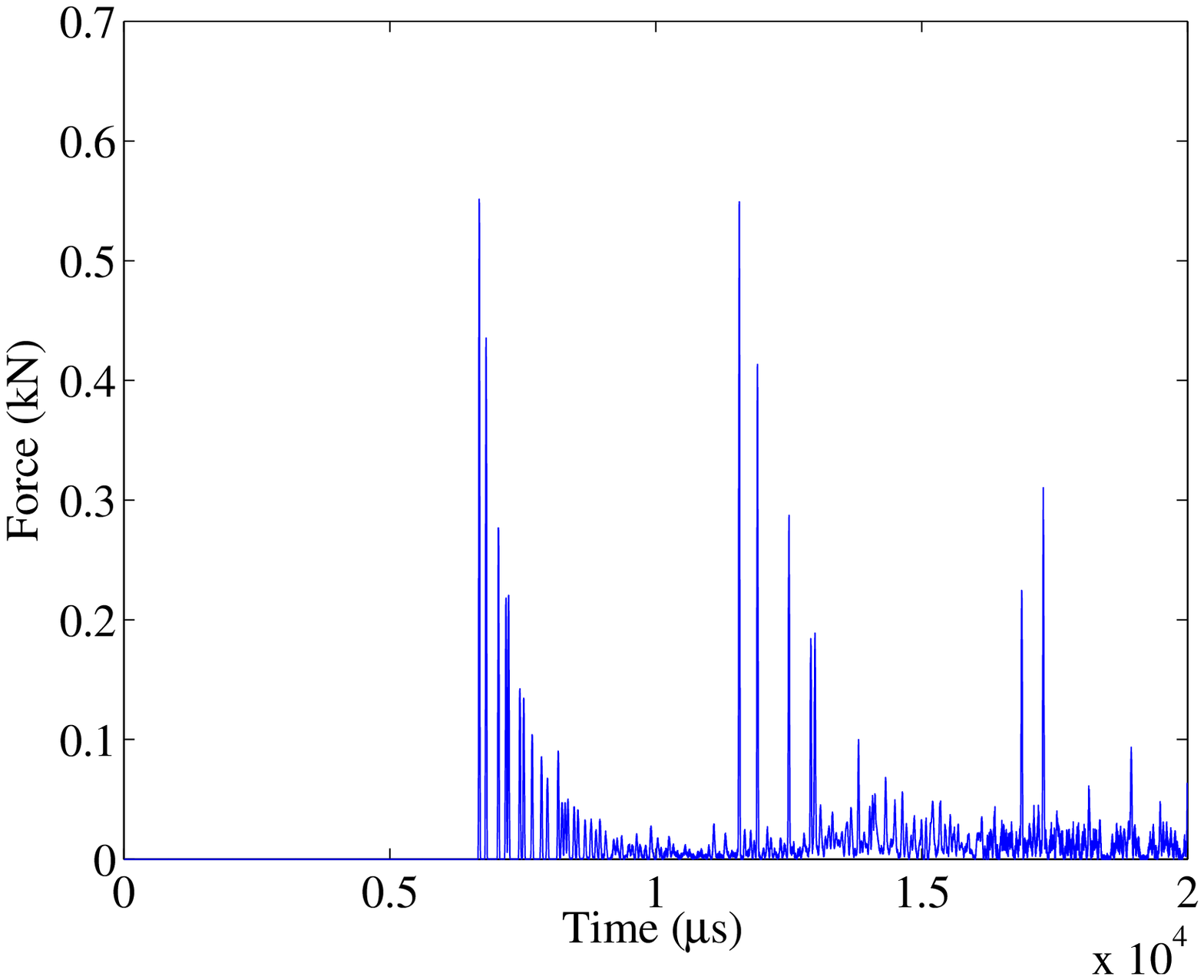,angle=0,width=44mm}       \end{array}$
     \caption{\scriptsize{(Left) Density plots (colored by force, with larger values given by darker shading) for (top row) decorated chain of Fig.~\ref{dec20_force} and for decorated chains with a single impurity in each 19-particle cell between beads 501 and 1000 (second row), 801 and 1000 (third row), and 951 and 1000 (bottom row).  (Right) Force (in kN) versus time plots for particle 1500 in each of these chains.}} 
     \label{spacetime}
  \end{center}
\end{figure}

Using long-wavelength asymptotics, one can obtain a nonlinear partial differential equation (PDE) description of the decorated chain in the continuum limit \cite{nesterenko1,dimer,dimerlong}.  This PDE has known compact solitary wave solutions, which we illustrate in the top panels of Fig.~\ref{spacetime}.  However, adding even a small number of impurities to the system can change things completely (introducing some  pulse disintegration, pulse reflection, and thermalization), although one still obtains a basically localized pulse.  The impurities we consider here consist of particles of radius 5 mm and mass 2.31 g, so that they are much larger and heavier than the other beads in the chain.  Throughout the region of the chain that has impurities, we place one of them every 19 particles, giving a cell length that is the same as that in the periodically repeated size-optimized chain of Fig.~\ref{opt20_force}.  The second through fourth rows of Fig.~\ref{spacetime} contain regions of different lengths that contain these periodic impurities.  In each case, the last impurity is placed before bead 1000.  The first impurity is in particle 501 in the second row of Fig.~\ref{spacetime}  The third row of the figure is for a chain with impurities every 19 beads starting from bead 801, and the bottom row is for a chain with impurities every 19 beads starting from bead 951 (so that there are three impurities in total--at beads 970 and 989--in this last example).  As shown in these plots, the insertion of such heavy impurities leads to partial reflections of the wave, some thermalization, significant pulse disintegration, and even to a bit of trapping (see the bottom left panel).  Also observe in the bottom row that we have induced delays in the wave reflection.  By tuning the material properties carefully, one can perhaps optimize the properties of such wave trapping so that they can be exploited in applications.  Moreover, these numerical experiments also illustrate the much more complicated series of secondary pulse emission and partial wave reflections that occur in the periodic sequence of optimized/randomized cells in Ref.~\ref{optspacetime}.

The study of soliton-like pulses in perturbed uniform Hertzian chains was reported earlier \cite{sen98, hong02}, suggesting their use as possible systems to detect buried impurities via the analysis of back-scattered signals. Our results show that similar phenomena can also be observed in the ``quasi-disordered" systems; as shown in Fig.~\ref{spacetime}, the mass and position of the defects in the chain detectably shift the reflection and the radiation.

It would be interesting to extend this type of discussion by considering increasingly disordered configurations, such as systems with quasiperiodic arrangements of cells (with various lengths and component particles) rather than periodic ones.  Some preliminary research in this direction (using, for example, arrangements that follow Fibonacci sequences) has appeared recently in the literature in order to study Anderson localization in atomic chains \cite{fib1,fib2}.  It would similarly be interesting to consider systems with random arrangements of cells.

\subsection{Material optimization}\label{material}

In Ref.~\cite{dar06}, Daraio, et al. investigated the optimization of a composite granular protector consisting of 22 stainless steel beads and 10 PTFE beads (see Table~\ref{tab:Mat1}) with a uniform radius of 2.38 mm. The authors examined different design solutions, based on material distribution, using both numerical and laboratory experiments. The protector they considered was impacted by an $\mbox{Al}_2 \mbox{O}_3$ striker (0.47 g) with initial velocity of 0.44 m/s and was initially precompressed by a static force of 2.38 N.  Using piezosensors embedded in selected particles, they obtained laboratory measurements of force versus time profiles in several beads and compared them against numerical predictions.  Here we report analogous experiments to confirm the results obtained from the optimization analysis. We used a four-garolite-rod stand as the holder for the beads and assembled sensors as described in Refs.~\cite{dimer,dimerlong}.  We selected a steel particle (0.45 g) as the striker and recorded the traveling signal with a TKTDS 2024 oscilloscope (Tektronix, Inc.). The sensors (PiezoSystems, Inc.) were calibrated by conservation of linear momentum. The 1.86 N static precompression included the preloading of the topmost particle with about 190 g of symmetrically suspended masses. 

\begin{figure}[tbp]
    \centerline{\includegraphics[angle=0,width=80mm]{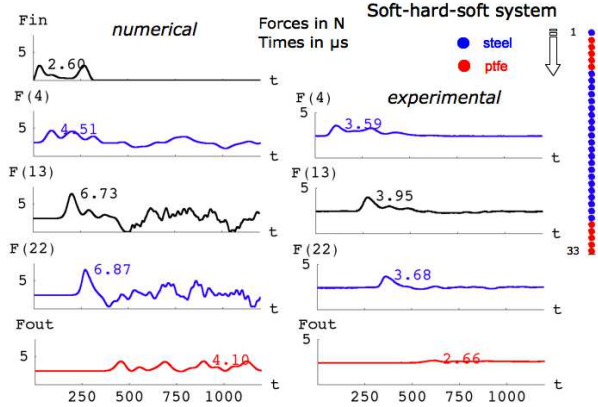}}
    \caption{\scriptsize{(Color online) Force versus time plots in a soft-hard-soft granular chain.  (The applied precompression was added to the force profiles experimentally recorded though piezosensors.)}}\label{FIG2_force}
\end{figure}

Figure~\ref{FIG2_force} shows some numerical and experimental force recordings in a soft-hard-soft configuration (see Fig.~2 of Ref.~\cite{dar06}) with two sequences of five PTFE beads at both the top and bottom of the chain.  This system had the minimum value of $F_{out}$ of all of the configurations considered in \cite{dar06}.  (A different hard-soft-hard-soft-hard configuration minimized $F_{out} /$ $F_{in}$ ratio, as shown in Fig.~3 of Ref.~\cite{dar06}.)  Observe the good qualitative agreement between numerical and experimental results over the initial phase of the pulse propagation.  The dynamics of the experiments and numerics subsequently deviate from each other, as the laboratory tests are affected by dissipative effects (not included in the numerics). The latter can arise from effects such as friction, inelastic collisions, viscous drag, etc.  The experimental traveling wave is thus progressively damped as it travels through the chain, resulting in an even better protector. 

\begin{figure}[tbp]
    \centerline{\includegraphics[angle=0,width=80mm]{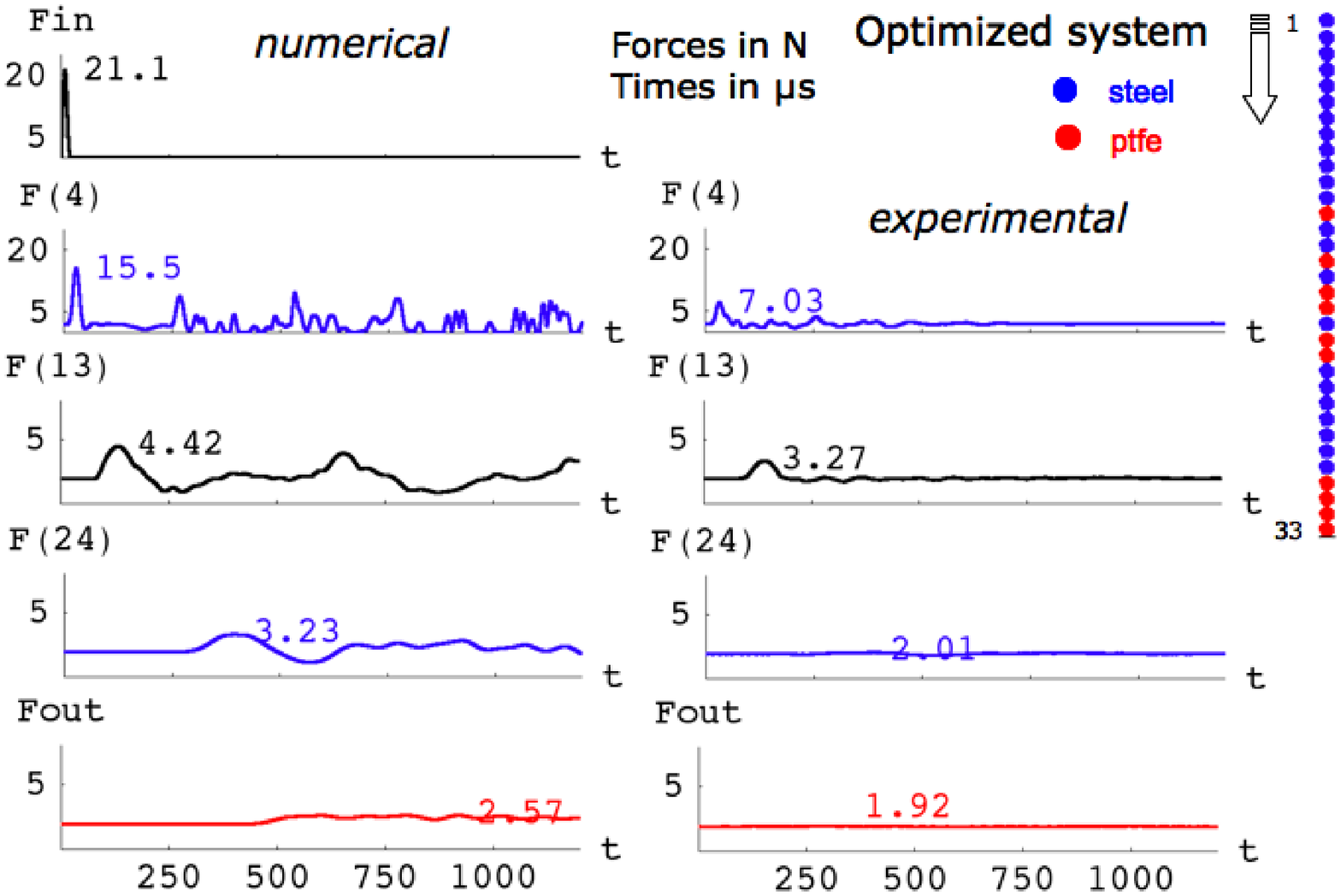}}
    \caption{\scriptsize{(Color online) Force versus time plots in the material-optimized system.  (The applied precompression was added to the force profiles experimentally recorded though piezosensors.)}}\label{opt33_force}
\end{figure}

\begin{figure}[htbp]
  \begin{center}
    \setlength{\unitlength}{1mm}
    $\begin{array}{cc}
    \epsfig{file=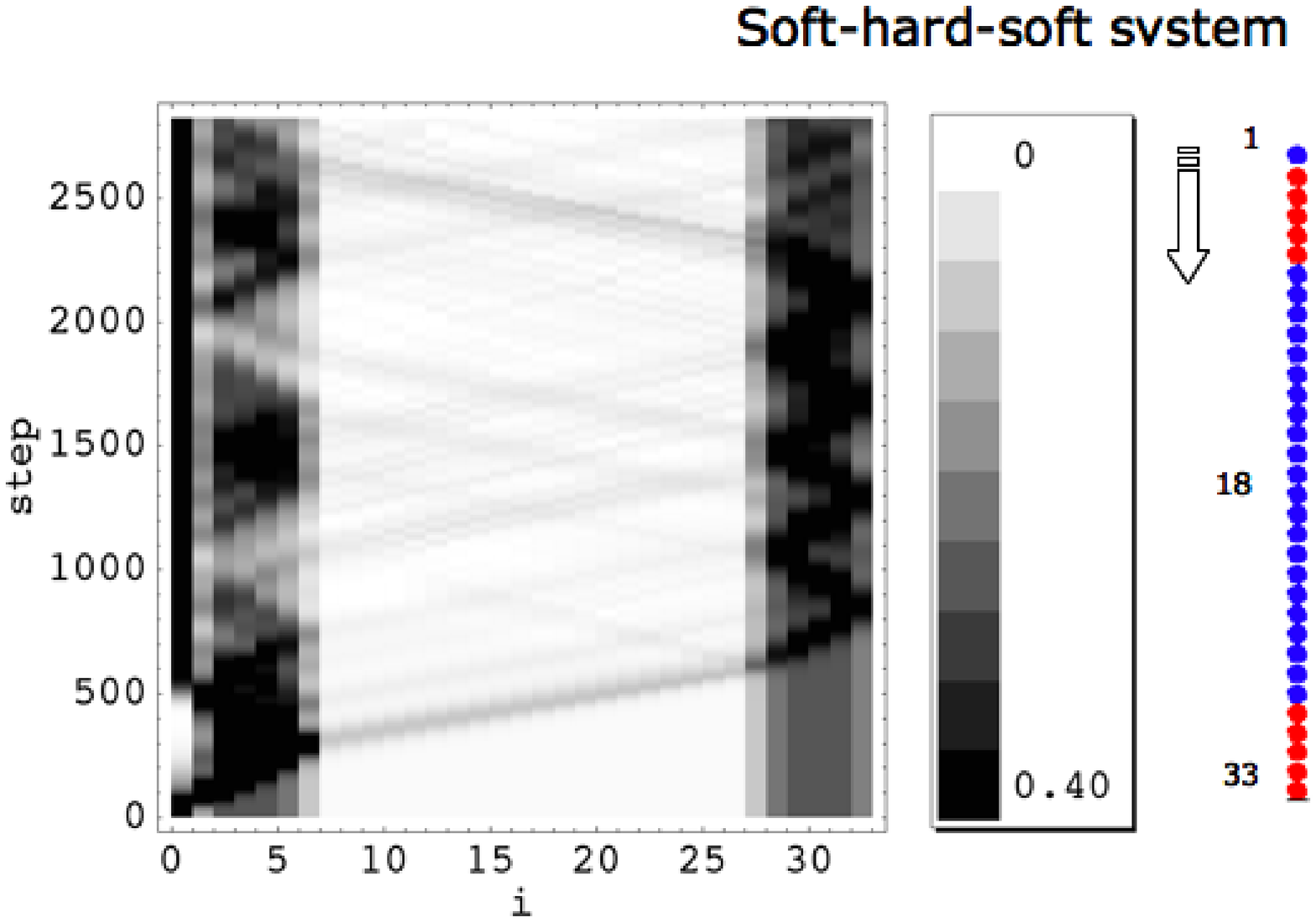,angle=0,width=44mm} &
    \epsfig{file=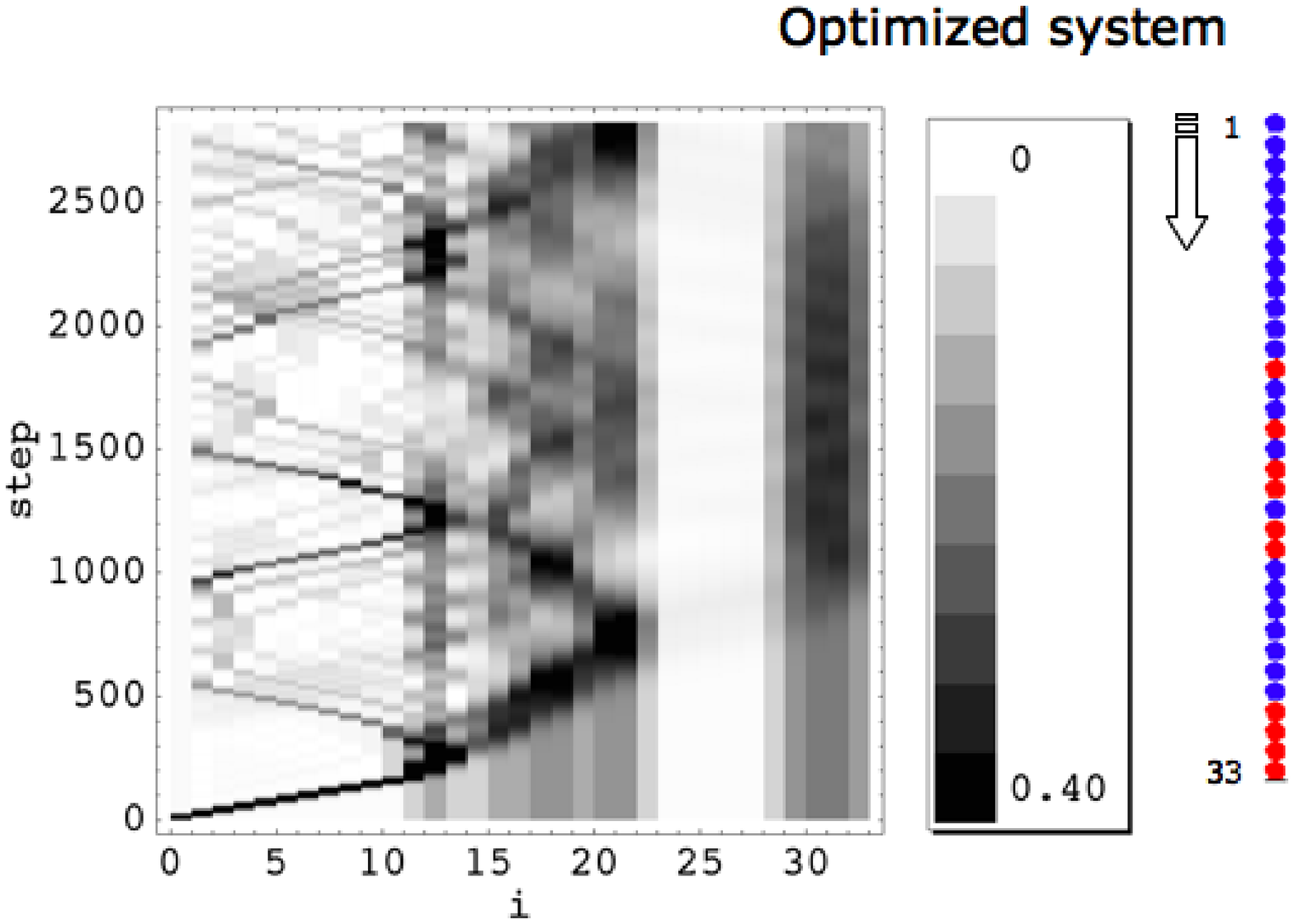,angle=0,width=44mm}       \end{array}$
     \caption{\scriptsize{(Color online) Density plots of particle energies normalized to unity.  Horizontal axes are labeled according to particle site and vertical axes give the time step.}}
     \label{material_density}
  \end{center}
\end{figure}

We carried out a material optimization by introducing $32$ genes $x_i$ related to the material identification of the individual beads (not including the striker). They are defined such that $x_i \in [0,0.5]$ implies that the $(i+1)$th bead is made of PTFE, whereas $x_i \in (0.5,1]$ implies that the same bead is instead made of stainless steel.  (We assumed that the striker was always made of steel.)  We constrained the total number of PTFE beads to be equal to 10 through a penalty technique.  We also introduced an additional gene (so that the total number of genes $M$ is equal to 33) related to the intensity of the preload, allowing the static precompression force $F_0$ to vary continuously within the interval  [0, 2.38] N ($F_0 = 2.38x_{33}$ N).  The optimized system, obtained after about 110 BGA generations, is shown in Fig.~\ref{opt33_force} together with the corresponding numerical and experimental force plots.  As in the previous examples, observe that the material-optimized system contains soft beads near the wall, hard beads near the end impacted by the striker, and alternating hard and soft beads in the central section of the chain.  We computed the optimal precompression force to be about 1.86 N.  Both the numerical and the experimental force plots of Fig.~\ref{opt33_force} show that the leading solitary wave first decomposes into a train of small pulses and subsequently mutates into an extended (long-wavelength), small-amplitude wave. The density plots of Fig.~\ref{material_density} illustrate the mechanisms of wave disintegration and reflection characterizing the dynamics of these systems. The experimental results presented in this article and in Ref.~\cite{dar06} confirm the enhanced performance of the BGA-optimized system (minimum $F_{out}$), as compared to all the other examined protectors.

\nobreak

\nobreak

\subsection{Long composite protector}\label{hongs}

\begin{figure}[tbp]
    \centerline{\includegraphics[angle=0,width=80mm]{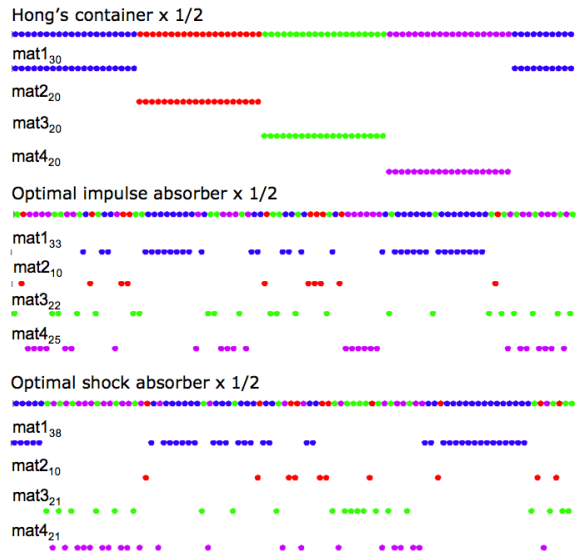}}
    \caption{\scriptsize{(Color online) Hong's container and optimized composite long chains (showing half of the chain; the other half is obtained by reflection symmetry).}}
   \label{hong_containers}
\end{figure}

In a recent paper \cite{hong05}, Hong investigated a long 1D composite granular protector (or ``energy container") consisting of nine 20-bead sections.  The beads in this chain were made of four different materials with varying particle mass $m$, contact stiffness $\alpha$, and contact exponent $n$ (see the parameter values in Table~\ref{tab:Mat2}). Hong used abstract units, introducing factors of $10^{-5}$ m, $2.36 \times 10^{-5}$ kg, and $1.0102 \times 10^{-3}$ s to convert the adopted units of length, mass, and time, respectively, into real units. The terminal and central sections of the protector are composed of a (linear) material (``material 1") characterized by a contact exponent $n=1$ and mass $m=2$. The inner sections are composed of beads (made of nonlinear materials) that have different masses and contact exponents greater than 1 (materials 2, 3, and 4).  This is used to simulated sharp contact surfaces and rough materials such as sand (see Fig.~\ref{hong_containers}).  The container has a reflection symmetry about its center and the initial distance between particle centers of mass is uniformly equal to $200$ along its body.  

Hong analyzed the behavior of this container (and variations thereof) by considering its dynamics after the impact of a striker of mass $m=100$ traveling with speed $v=10$.  He ran numerical simulations of the chain dynamics, employing a hard-sphere model and introducing lateral constraints through additional beads consisting of very heavy grains ($m=100$).  He found a universal power-law scaling for how long it took the energy to leak from the protector to the lateral sides.  That is, the energy remaining in the protector is given by $E_R = A t^{-\gamma}$, where $A$ is a constant that depends on the protector construction, $t$ is the time, and $\gamma$ is a constant (that Hong estimated in Ref.~\cite{hong05} to be about $0.7055$) independent of the protector construction.

We carried out a joint topology-material optimization of Hong's container (shown in the top panel of Fig. \ref{hong_containers}), introducing $M = 90$ genes $x_i$ that characterize the material identification of each individual bead.  The conditions $x_i \in [0,0.25]$, $x_i \in (0.25,0.5]$, $x_i \in (0.5,0.75]$, and $ x_i \in (0.75,1]$ respectively imply that the $i$th bead of one half of the protector is composed of material 1, 2, 3, and 4.  We enforced the symmetry with respect to the center of the chain by suitably relating the material identification numbers of the 90 grains of the second half to those of the grains in the first half.  We did not enforce any constraints on the numbers of beads of the different materials. Due to the central symmetry constraint, the optimized protector is not allowed to have different constructions near the impacted and the constrained ends, in contrast to the protectors we examined in the previous sections.

\nobreak

\subsubsection{Impulsive loading}

\begin{figure}[tbp]
    \centerline{\includegraphics[angle=0,width=80mm]{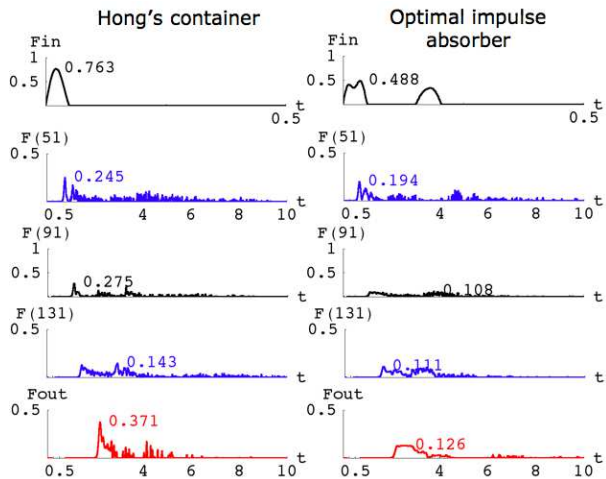}}
    \caption{\scriptsize{(Color online) Force versus time plots in a long composite chain subject to impulsive loading (force values divided by 1000).}}\label{impulse_force}
\end{figure}

\begin{figure}[htbp]
  \begin{center}
    \setlength{\unitlength}{1mm}
    $\begin{array}{cc}
    \epsfig{file=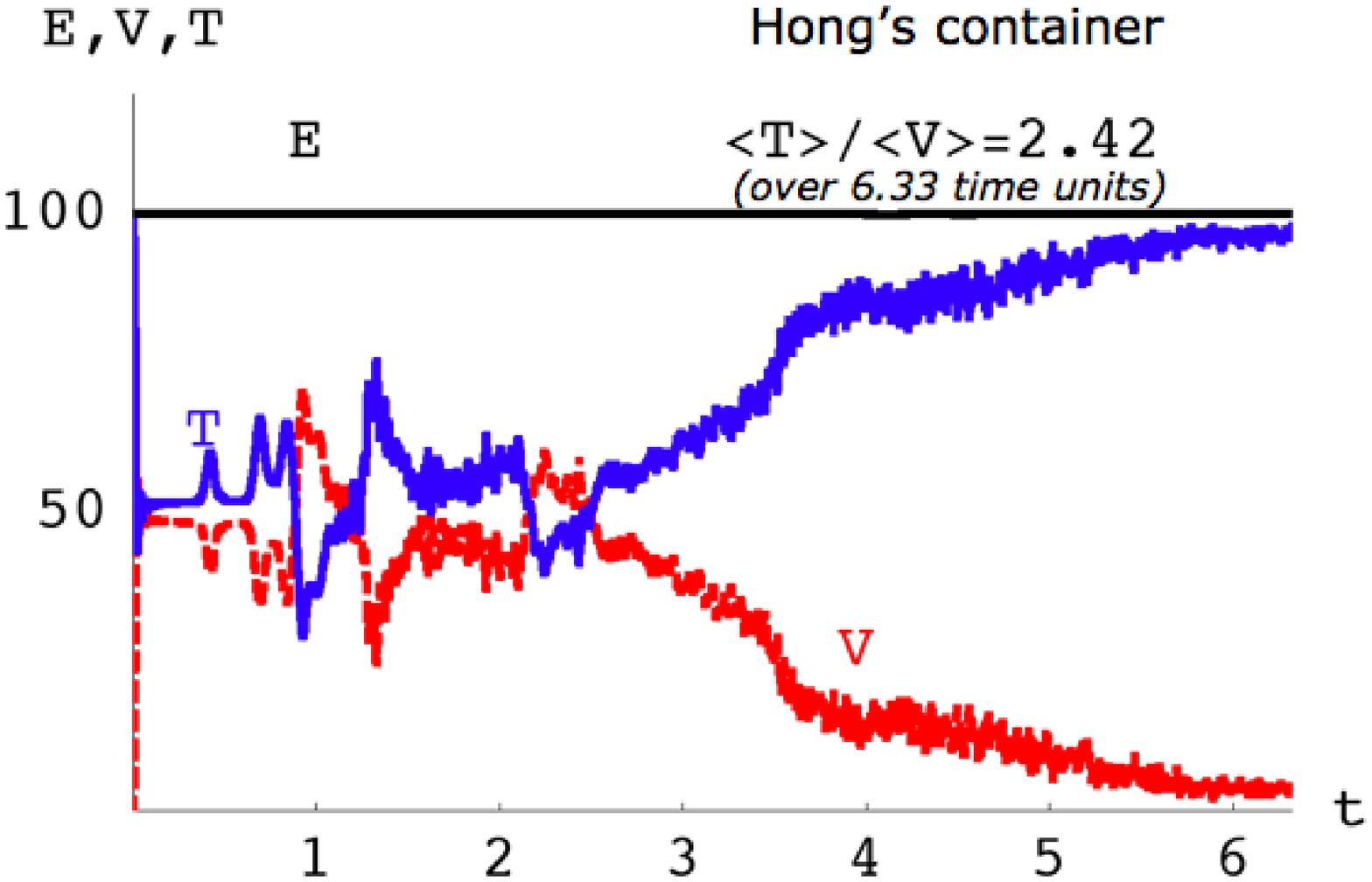,angle=0,width=44mm} &
    \epsfig{file=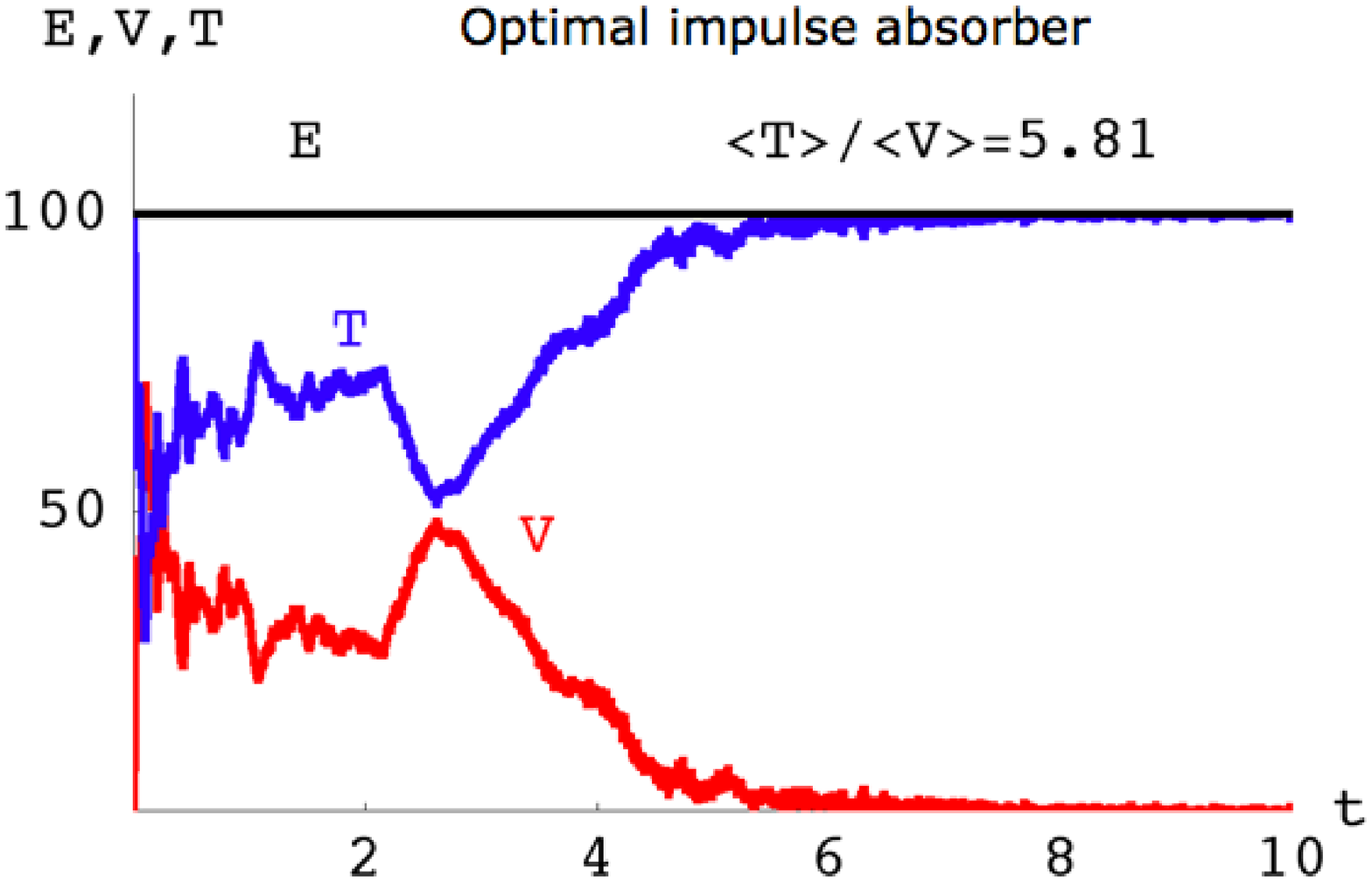,angle=0,width=44mm}       \end{array}$
     \caption{\scriptsize{(Color online) Energy versus time plots in Hong's container and the optimized chain under impulsive loading.}}
     \label{impulse_energy}
  \end{center}
\end{figure}

\begin{figure}[htbp]
  \begin{center}
    \setlength{\unitlength}{1mm}
    $\begin{array}{cc}
    \epsfig{file=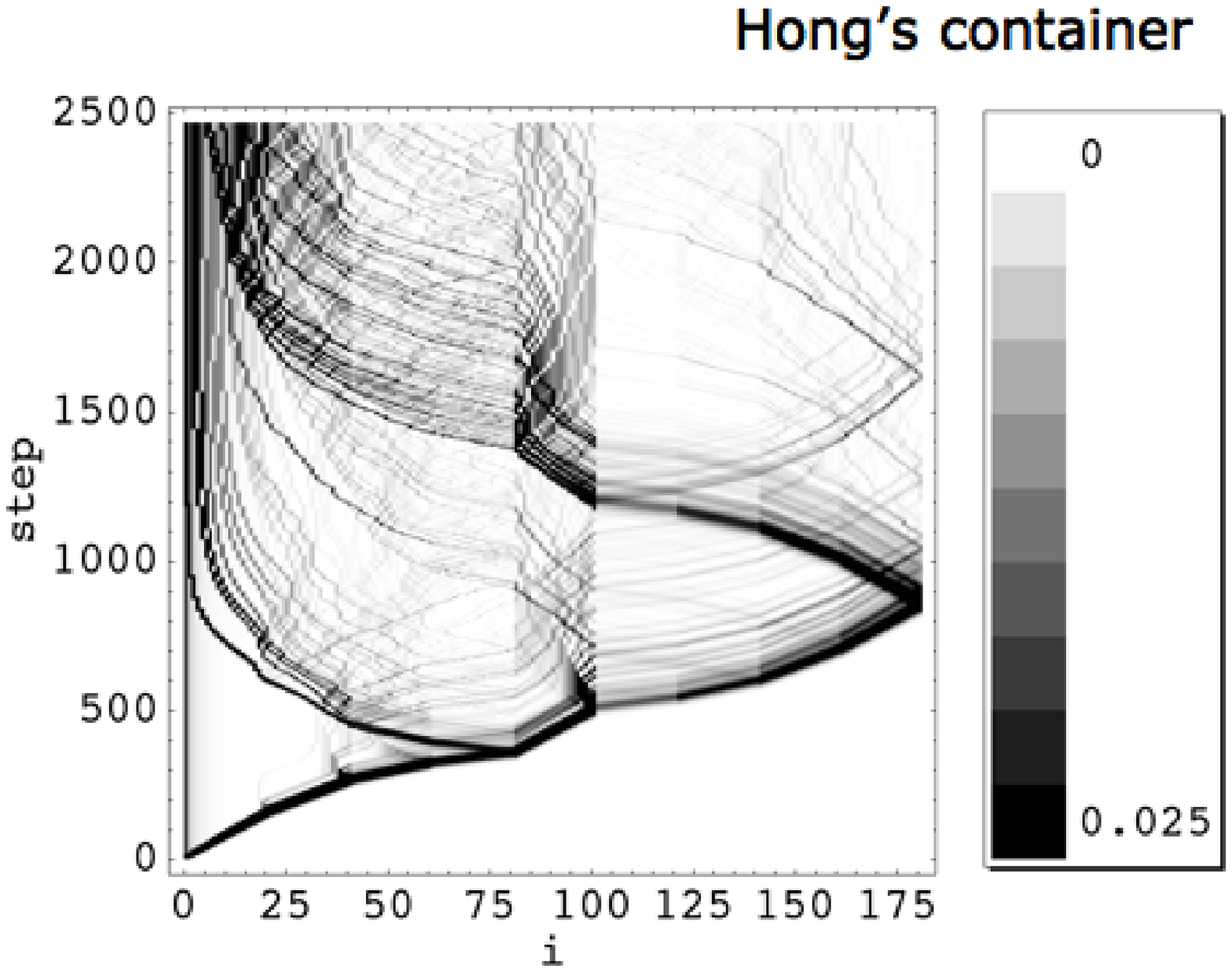,angle=0,width=44mm} &
    \epsfig{file=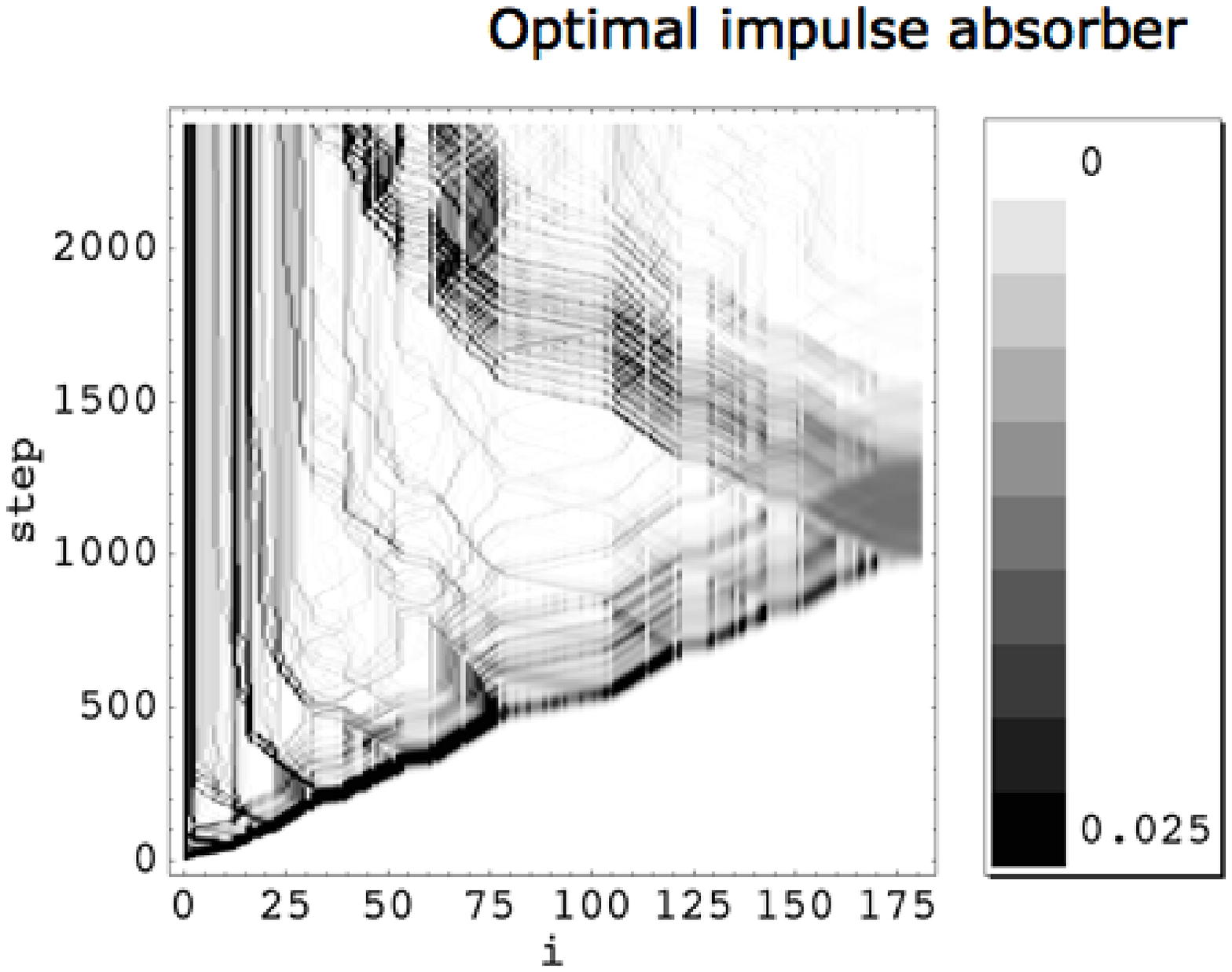,angle=0,width=44mm}       \end{array}$
     \caption{\scriptsize{Density plots of particle energies normalized to unity.  Horizontal axes are labeled according to particle site and vertical axes give the time step.}}
     \label{impulse_density}
  \end{center}
\end{figure}

\begin{figure}[tbp]
    \centerline{\includegraphics[angle=0,width=60mm]{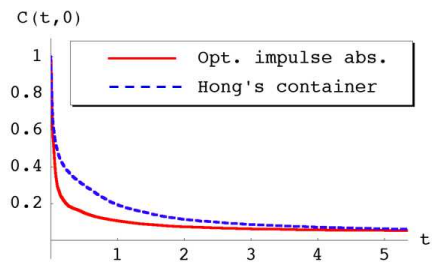}}
    \caption{\scriptsize{(Color online) Energy correlation function versus time for the examined systems.}}\label{impulse_corr}
\end{figure}

In our first optimization procedure, we considered the impact of an $m=2$ striker (material 1) traveling with speed $v=10$. We show the corresponding optimized protector, which we obtained after about 400 BGA generations, in Fig.~\ref{hong_containers}. This \textit{optimal impulse absorber} has nonlinear beads near its extremities and in its center and sequences of linear beads in its remaining sections (a few of them are also near the center of the half chain).  We show the corresponding force plots and energy profiles (as well as the ones for Hong's container) in Fig.~\ref{impulse_force} and \ref{impulse_energy}, respectively.  Observe that the optimized scheme transmits to the wall a maximum force (126) that is about three times smaller than that transmitted by the basic scheme (371).  As in the previous examples, the initial pulse is progressively disintegrated, reflected, and transformed into an extended wave within the optimized system (as confirmed also by the density plots of Fig.~\ref{impulse_density}). Hong's container instead shows wave reflection only when the incident wave crosses the central section of the system. The time histories of the energy correlation function, depicted in Fig.~\ref{impulse_corr}, highlight the fact that the optimized system spreads out energy (i.e., thermalizes) on a faster time scale than Hong's container.

\nobreak

\nobreak

\subsubsection{Shock-type loading}

\begin{figure}[tbp]
    \centerline{\includegraphics[angle=0,width=80mm]{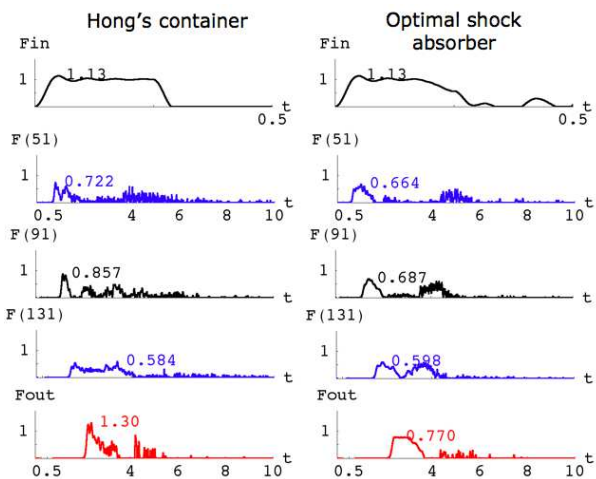}}
    \caption{\scriptsize{(Color online) Force versus time plots in a long composite chain subject to shock-type loading (forces values divided by 1000).}}\label{shock_force}
\end{figure}

\begin{figure}[htbp]
  \begin{center}
    \setlength{\unitlength}{1mm}
    $\begin{array}{cc}
    \epsfig{file=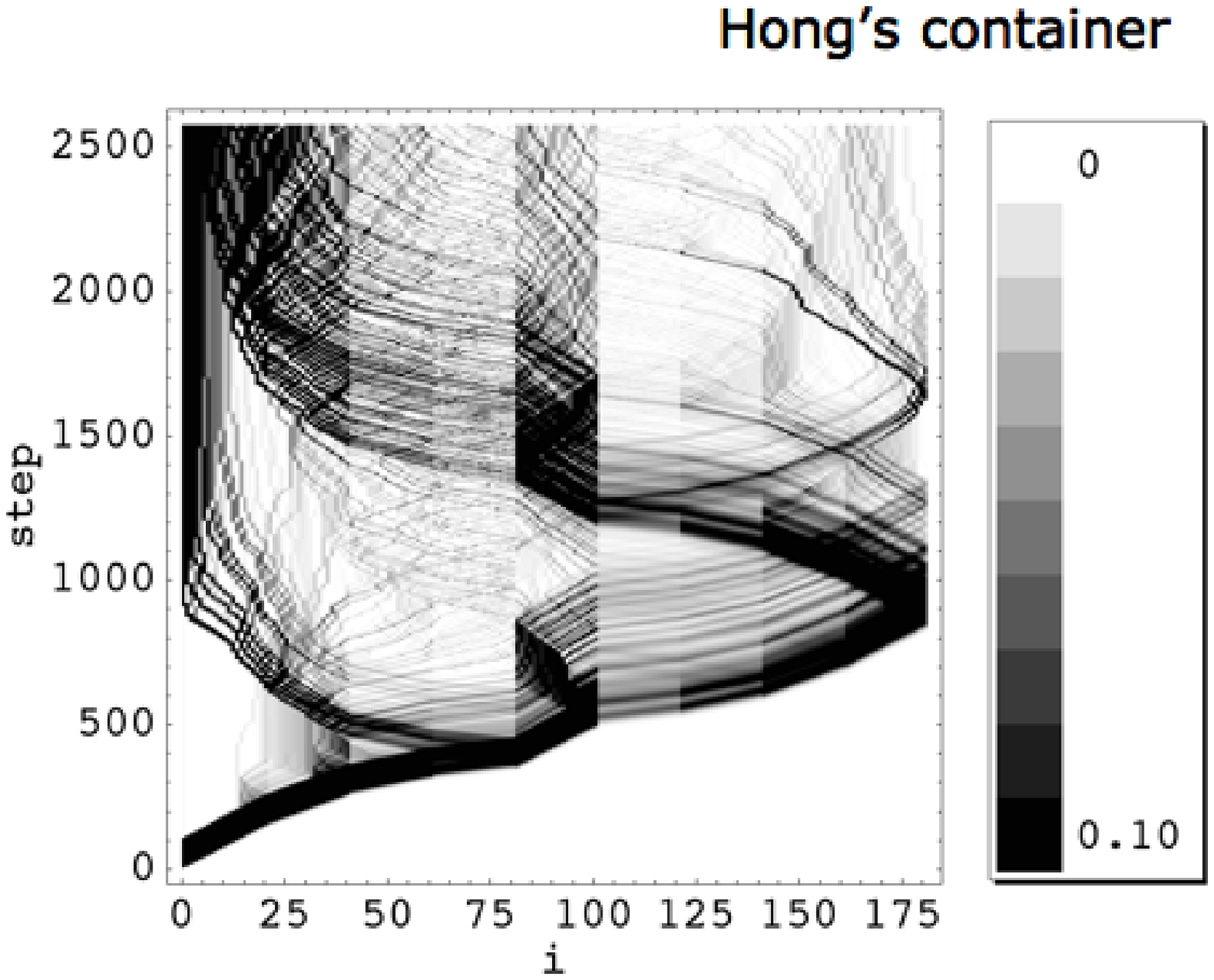,angle=0,width=44mm} &
    \epsfig{file=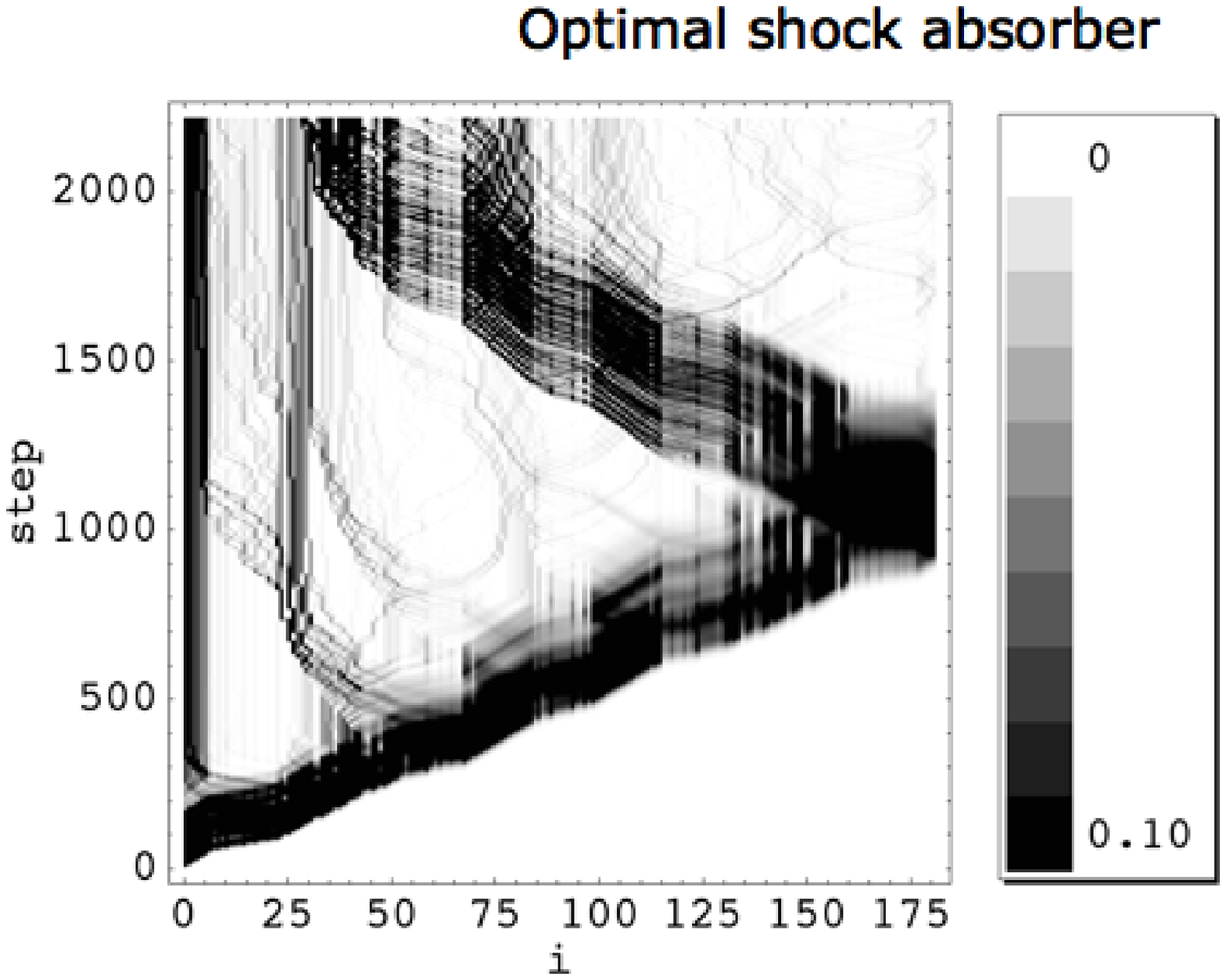,angle=0,width=44mm}       \end{array}$
     \caption{\scriptsize{Density plots of particle energies normalized to unity.  Horizontal axes are labeled according to particle site and vertical axes give the time step.}}
     \label{shock_density}
  \end{center}
\end{figure}

We also carried out an optimization procedure using a shock-type force profile with constant intensity $F=1000$ on the first bead (composed of material 1), external to the container, for a time equal to 0.25. We show the corresponding optimal container, which we obtained after about 200 BGA generations, in Fig.~\ref{hong_containers}.  The force-time plots of this \textit{optimal shock absorber} and those of the basic Hong scheme, with the shock-type loading, are shown in Fig.~\ref{shock_force}.  Observe that the basic protector transmits to the wall a peak force (1300 units) larger than the applied shock, whereas the optimized system is able to reduce the shock at the wall up to 770 units (a 23$\%$ reduction).  Note additionally that the optimal shock absorber is characterized by heavy, linear (material 1) grains near the extremities (see Fig.~\ref{hong_containers}).  This is likely due to the symmetry constraint discussed above.  Figure~\ref{shock_force} also shows that the input shock gets weakened when traveling along the optimal protector.  We show the density plots of particle energies in these two systems in Fig.~\ref{shock_density}.

\nobreak

\section{Conclusions} \label{conclusions}

In summary, we used an evolutionary algorithm to investigate the optimal design of composite granular protectors using one-dimensional chains of beads composed of materials of various size, mass, and stiffness.  Identifying the maximum force $F_{out}$ transmitted from the protector to a ``wall" that represents the body to be protected as a fitness function, we optimized the {topology} (arrangement), {size}, and {material} of the beads in the chain in order to minimize $F_{out}$.  We considered several examples that were investigated recently in the literature, including stepped two sonic vacua, tapered chains, decorated chains, and a recent configuration due to Hong.  

The optimization procedure, driven by a Breeder Genetic Algorithm, produced (optimally) randomized/disordered systems, along which the incident waves were disintegrated and reflected, exhibiting marked thermalization.  Additionally, the solitary pulses traveling to the wall combined to form extended (long-wavelength), small-amplitude waves.  In the absence of enforced central symmetry, the optimal configurations had soft/light beads near the wall, hard/heavy beads near the loaded end, and alternating hard/heavy and soft/light beads in the remaining part of the chain.  In the presence of central symmetry, we instead obtained an optimal configuration that had light, nonlinear beads toward the ends in the case of impulsive loading and one that had heavy, linear beads toward the ends in the case of shock-type loading.

The present research paves the way for many interesting developments, as our approach can be generalized to numerous situations.  First, the techniques we employed can be applied to more intricate experimental configurations--including two-dimensional systems, three-dimensional systems,  systems composed of ensembles of particles with non-spherical geometries or even layered materials.  Second, one can incorporate additional physical effects, such as dissipation and more complicated contact mechanics.  Third, one can generalize the methods themselves by, for example, adopting continuous optimization techniques such as the material distribution method \cite{ben03} and the formulation of multiple-scale approaches that involve scale-dependent interaction forces.

\section*{Acknowledgements}

We thank Dr. Antonio Della Cioppa from the Department of Information and Electrical Engineering of the University of Salerno for providing the BGA code.  F.F.\ greatly acknowledges the support of the Italian MIUR through the 2007 grant ``Energetic Methods in Fracture Mechanics and Biomechanics."  FF and MAP thank the Graduate Aeronautics laboratory at Caltech (GALCIT) for hospitality during their visits, and CD acknowledges support from Caltech startup funds.


\end{document}